\documentclass[10pt, letter, onecolumn]{arxiv}

\usepackage{kantlipsum, lipsum}
\usepackage{dm-colors}
\usepackage{amsmath}
\usepackage{pstricks, pst-node}
\usepackage{verbatim}
\usepackage{multirow}
\usepackage{scalerel}
\usepackage{booktabs}
\usepackage{enumitem}
\usepackage{xspace}
\usepackage{bm}
\usepackage{bbm}
\usepackage{mathtools}
\usepackage{soul}
\usepackage{epsfig}
\usepackage{graphicx}
\usepackage{amssymb}
\usepackage{colortbl}
\usepackage{csquotes}
\usepackage{setspace}
\usepackage{colortbl}
\usepackage{bbding}
\usepackage{threeparttable}
\usepackage{tabularx,ragged2e}
\usepackage{placeins}
\usepackage{bbding}

\definecolor{darkpastelgreen}{rgb}{0.13, 0.55, 0.13}

\definecolor{darkpastelred}{rgb}{0.55, 0.13, 0.13}

\usepackage[hang,flushmargin]{footmisc}

\usepackage{nameref}
\usepackage{varioref}
\usepackage{amssymb}
\usepackage{pifont}

\usepackage[pagebackref=false,breaklinks=false,
            colorlinks=true,bookmarks=true,citecolor=ourdarkblue,
            urlcolor=ourdarkblue,linkcolor=ourdarkblue]{hyperref}
\usepackage[noabbrev,capitalize]{cleveref}
\usepackage{etoc}

\usepackage[numbers,sort&compress]{natbib} 

\graphicspath{{figures/}}

\title{\Large{Cardiac-CLIP: A Vision-Language Foundation Model for \\ 3D Cardiac CT Images}}

\vspace{1cm}
\author[1,$\ast$]{Yutao Hu} 
\author[1,$\ast$]{Ying Zheng} 
\author[3,1]{Shumei Miao} 
\author[2]{Xiaolei Zhang} 
\author[1]{Jiahao Xia} 
\author[1]{Yaolei Qi} 
\author[1]{Yiyang Zhang} 
\author[16]{\\Yuting He} 
\author[5]{Qian Chen} 
\author[6]{Jing Ye} 
\author[7]{Hongyan Qiao} 
\author[8]{Xiuhua Hu} 
\author[9]{Lei Xu} 
\author[10]{Jiayin Zhang} 
\author[11]{Hui Liu} 
\author[12]{\\ Minwen Zheng} 
\author[13]{Yining Wang} 
\author[14]{Daimin Zhang} 
\author[15]{Ji Zhang} 
\author[4]{Wenqi Shao}
\author[3]{Yun Liu} 
\author[2,$\dag$]{\\ Longjiang Zhang} 
\author[1,$\dag$]{Guanyu Yang}

\makeatletter
\renewcommand\AB@affilsepx{\protect\\}
\makeatother

\affil[1]{\normalsize School of Computer Science and Engineering, Southeast University, Nanjing, China} 
\affil[2]{\normalsize Department of Radiology, Jinling Hospital, Affiliated Hospital of Medical School, Nanjing University, Nanjing, China}
\affil[3]{\normalsize The First Affiliated Hospital of Nanjing
Medical University, Nanjing, China}
\affil[4]{\normalsize Shanghai AI Laboratory, Shanghai, China}
\affil[5]{\normalsize Department of Radiology, Nanjing First Hospital, Nanjing Medical University, Nanjing, China}
\affil[6]{\normalsize Radiology Department, Northern Jiangsu People's Hospital, Yangzhou, China}
\affil[7]{\normalsize Department of Medical Imaging, the Affiliated Hospital of Jiangnan University, Wuxi, China}
\affil[8]{\normalsize Sir Run Run Shaw Hospital, Zhejiang University School of Medicine, Hangzhou, China}
\affil[9]{\normalsize Department of Radiology, Beijing Anzhen Hospital, Capital Medical University, Beijing, China.}
\affil[10]{\normalsize Department of Radiology, Shanghai General Hospital, Shanghai Jiao Tong University School of Medicine, Shanghai, China.}
\affil[11]{\normalsize Department of Radiology, Guangdong Provincial People's Hospital, Guangzhou, China}
\affil[12]{\normalsize Department of Radiology, Xijing Hospital, Air Force Medical University, Xi'an, China.}
\affil[13]{\normalsize Department of Radiology, Peking Union Medical College Hospital, Chinese Academy of Medical Sciences and Peking Union Medical College, Beijing, China}
\affil[14]{\normalsize Department of Cardiology, Sir Run Run Hospital, Nanjing Medical University, Nanjing, China.}
\affil[15]{\normalsize Department of Radiology, Taizhou People's Hospital, Taizhou, China}
\affil[16]{\normalsize Case Western Reserve University, Cleveland, United States}

\renewcommand{\correspondingauthor}[1]{$\ast$~Equal contributions. \\ $\dag$~Corresponding author. }

\begin{document}

\begin{abstract}
Foundation models have demonstrated remarkable potential in medical domain. However, their application to complex cardiovascular diagnostics remains underexplored. In this paper, we present Cardiac-CLIP, a multi-modal foundation model designed for 3D cardiac CT images. Cardiac-CLIP is developed through a two-stage pre-training strategy. The first stage employs a 3D masked autoencoder (MAE) to perform self-supervised representation learning from large-scale unlabeled volumetric data, enabling the visual encoder to capture rich anatomical and contextual features. In the second stage, contrastive learning is introduced to align visual and textual representations, facilitating cross-modal understanding. To support the pre-training, we collect 16641 real clinical CT scans, supplemented by 114k publicly available data. Meanwhile, we standardize free-text radiology reports into unified templates and construct the pathology vectors according to diagnostic attributes, based on which the soft-label matrix is generated to supervise the contrastive learning process. On the other hand, to comprehensively evaluate the effectiveness of Cardiac-CLIP, we collect 6,722 real-clinical data from 12 independent institutions, along with the open-source data to construct the evaluation dataset. Specifically, Cardiac-CLIP is comprehensively evaluated across multiple tasks, including cardiovascular abnormality classification, information retrieval and clinical analysis. Experimental results demonstrate that Cardiac-CLIP achieves state-of-the-art performance across various downstream tasks in both internal and external data. Particularly, Cardiac-CLIP exhibits great effectiveness in supporting complex clinical tasks such as the prospective prediction of acute coronary syndrome, which is notoriously difficult in real-world scenarios. As far as we know, our work is the first attempt to develop a cardiac foundation model based on such large-scale data. We hope our research could support clinical decision-making and facilitate further research in this area.

\end{abstract}

\maketitle

\section{Introduction}
Cardiovascular disease has emerged as a predominant global health threat, accounting for escalating health and economic burdens worldwide. Epidemiological data reveal a striking increase in cardiovascular disease mortality, rising from 12 million annual deaths in 1990 to more than 19 million in 2021, with both incidence and mortality rates exhibiting sustained upward trends \cite{1_1_2025_Heart_disease_statistics,1_2_China2020_2030}. The increasing prevalence of cardiovascular disease not only endangers population health but also places considerable strain on clinical and financial resources, underscoring the urgent need for effective strategies in prevention, diagnosis, and treatment. In the diagnosis of cardiovascular disease, cardiac CT, including coronary artery calcium computed tomography (CACCT) and coronary computed tomography angiography (CCTA) are two essential imaging modalities \cite{zeleznik2021deep, fairbairn2025implementation}. Therefore, to facilitate more accurate and efficient diagnosis of cardiovascular disease, precise interpretation of computed tomography (CT) image is critically important.

With the rapid advancement of artificial intelligence (AI) and computer vision, leveraging AI-based techniques to analyze cardiac CT images has received increasing attention. In recent years, deep learning have enabled efficient analysis of cardiac images with promising performance\cite{3_1_AI4CVD_care_review,3_2,3_3Yang2016,3_4Wolterink2016,3_5_NC_Zeleznik2021,3_6Eng2021,3_8Zreik2019,3_9_NC_Chao2021,zreik2019deep, van2020deep}, facilitating key clinical applications such as coronary segmentation \cite{qi2023dynamic}, plaque characterization \cite{denzinger2019coronary} and cardiovascular disease risk stratification \cite{dey2018integrated}. Despite their revolutionary impact, many challenges still remain to be addressed. Specifically, one of the most significant problem is the reliance on meticulously annotated data, which significantly increases the cost and effort required for network training \cite{tizhoosh2018artificial, wang2021annotation}. This issue is particularly pronounced in cardiac CT, where the pathological features of interest are often subtle and require expert-level annotation \cite{tolle2025real}. Moreover, models trained on narrowly defined disease categories frequently exhibit poor generalization capacity to unseen conditions \cite{guan2021domain}. Meanwhile, significant variations often exist in the distribution of CT imaging data acquired from different medical institutions, which necessitates repeated annotation and retraining to mitigate domain shifts. As a result, the limited generalization capability of existing models significantly restricts their practical applicability in real-world clinical settings.

Recently, vision-language foundation models \cite{4_2OrigCLIP, jia2021scaling, singh2022flava} pre-trained on large-scale paired image–text data have demonstrated strong generalization capabilities without the need for exhaustive manual labeling. These models exploit the natural co-occurrence of images and accompanying text to learn semantically meaningful associations via contrastive learning framework. Such a paradigm align particularly well with medical imaging workflows, where radiology reports are routinely generated alongside corresponding scans. Consequently, following this paradigm, several foundation models have been proposed in the medical field \cite{4_3_N_pathology_Xiang2025,4_6_NM_pathology_Lu2024,4_7_NM_pathology_Huang2023,4_8_KnowledgeEn_NC_X_Zhang2023,4_9_MAE_NC_X_Huang2024,4_11_NMI_X_Zhou2022,4_15_KnowledgeEn_X_Wu_2023_ICCV,4_16_prompt_soft_X_Lai2024,4_17_soft_X_Wang2022,4_12_NM_skin_Kim2024,4_13_NC_skin_Zhou2024,5_4_blankemeier2024merlin,5_5_Hamamci2024,5_6_bai2024m3d,5_1_soft_CHestCT_Cao2024}. Among these models, some aim for broad applicability across multiple modalities\cite{4_17_soft_X_Wang2022, lin2023pmc, 5_6_bai2024m3d}, which often comes at the expense of domain-specific knowledge. As a result, their utility in complex clinical scenarios is significantly limited. On the other hand, some approaches focus on specific imaging modalities or anatomies (\emph{e.g.,} pathology image\cite{4_3_N_pathology_Xiang2025,4_6_NM_pathology_Lu2024,4_7_NM_pathology_Huang2023}, chest X-ray \cite{4_8_KnowledgeEn_NC_X_Zhang2023,4_9_MAE_NC_X_Huang2024,4_11_NMI_X_Zhou2022,4_15_KnowledgeEn_X_Wu_2023_ICCV,4_16_prompt_soft_X_Lai2024,4_17_soft_X_Wang2022,8_2_M3AE_Chen2022}, chest CT\cite{5_5_Hamamci2024,5_1_soft_CHestCT_Cao2024}, dermatological image\cite{4_12_NM_skin_Kim2024,4_13_NC_skin_Zhou2024}, retinal scan \cite{4_14_N_retinal_Zhou2023}, abdominal CT\cite{5_4_blankemeier2024merlin}). Nevertheless, most existing medical vision–language models are designed for 2D slices \cite{4_17_soft_X_Wang2022, lin2023pmc, 4_15_KnowledgeEn_X_Wu_2023_ICCV}, with only a few efforts \cite{5_4_blankemeier2024merlin, 5_5_Hamamci2024, 5_6_bai2024m3d} extending to 3D medical data. Meanwhile, the acquisition of high-quality cardiac images is difficult, and no publicly available datasets currently exist to support large-scale foundation model development in this domain. As a result, there remains a notable absence of foundation models that are clinically applicable for cardiac CT.

\begin{figure}[htbp]
\centering
    \includegraphics[width=1.0\linewidth]{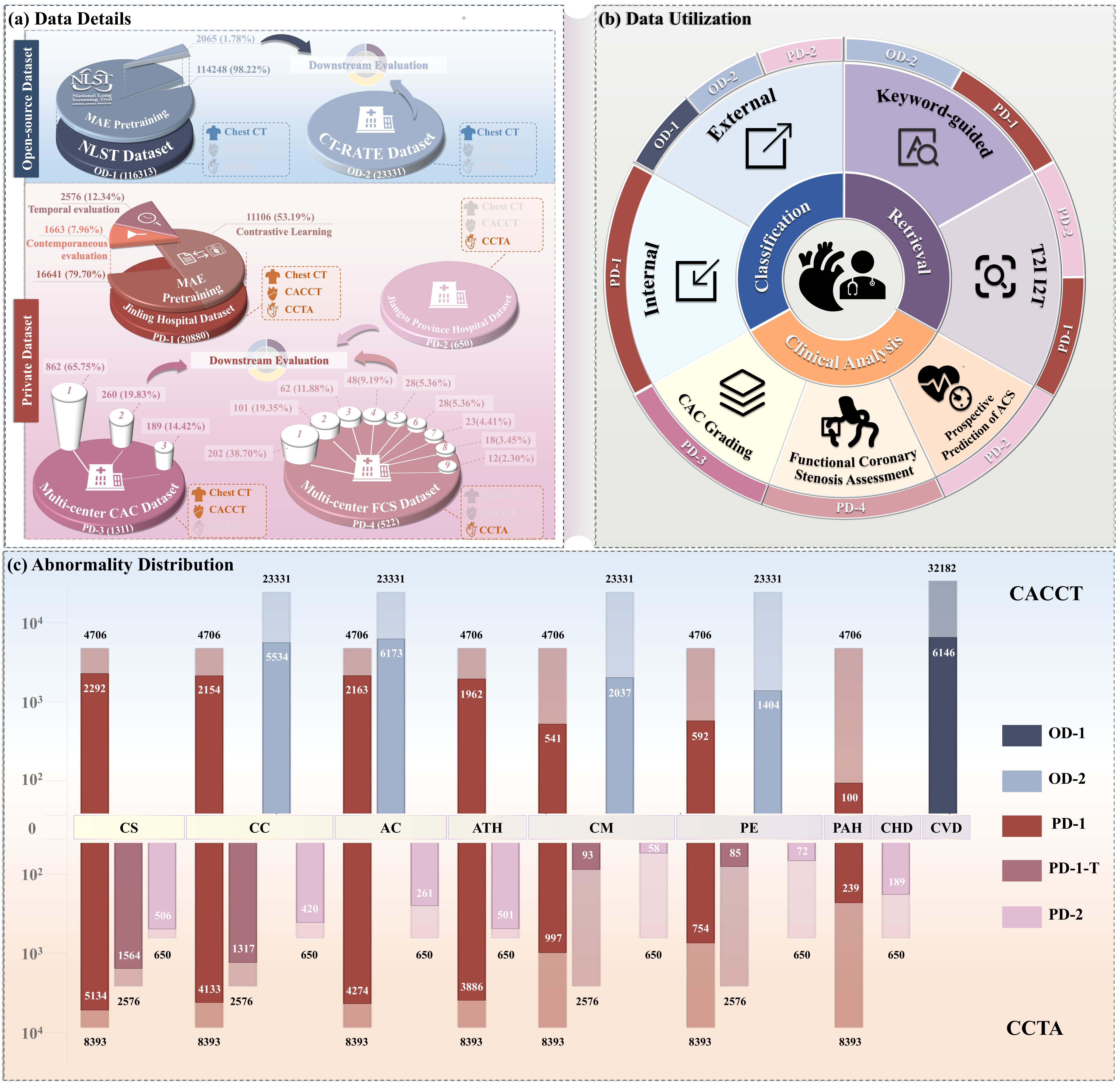}
\caption{Overview of the datasets utilized in this work. \textbf{(a)} Overview of the datasets utilized in this study. \textbf{(b)} The involved dataset of different downstream evaluation tasks. \textbf{(c)} Distribution and frequency of cardiovascular abnormalities present in each datasets. Based on these abnormalities, we perform classification experiments. Notably, the abbreviations in the figure are defined as follows: \textbf{CS}: Coronary Stenosis, \textbf{CC}: Coronary Calcification, \textbf{AC}: Aortic Calcification, \textbf{ATH}: Atherosclerosis, \textbf{CM}: Cardiomegaly, \textbf{PE}: Pericardial Effusion, \textbf{PAH}: Pulmonary Arterial Hypertension, \textbf{CHD}: Coronary Heart Disease, \textbf{CVD}: Cardiovascular Disease. Note that overlap may exist between abnormalities due to their extraction from radiology reports across different hospitals and public datasets. However, such overlaps only exist between different datasets. since the performance evaluation for each dataset is conducted independently, the overlaps do not affect the validity of our experimental results.}
\label{fig:data}
\end{figure}

To address the aforementioned limitations, we propose Cardiac-CLIP, a vision-language foundation model specifically designed for cardiac CT images. To facilitate robust multi-modal pre-training, we collect 11,106 real-world cardiac CT and 5,535 chest CT scans from Jinling Hospital, each paired with radiologist-authored diagnostic reports. To the best of our knowledge, this is the first attempt to develop a foundation model trained on such large-scale real-world clinical cardiac data. Additionally, to enrich the training data and promote the visual representation capacity of visual encoder, publicly available chest CT dataset, NLST \cite{7_NLSTdata_2011}, is also incorporated in the pre-training. Following the strategy in \cite{3_5_NC_Zeleznik2021}, all the chest CT images in NLST are pre-processed with cardiac region cropping. Overall, as shown in Fig.~\ref{fig:data}, there are totally 130,889 CT scans are employed in the training. To fully leverage these large-scale multi-modal cardiac data, as illustrated in Fig.~\ref{fig:framework}, we adopt a two-stage pre-training strategy for developing Cardiac-CLIP. In the first stage, masked autoencoder (MAE) \cite{8_MAE_He2022} is performed based on the entire training data, \emph{e.g.,} 130,899 CT scans from the NLST dataset and Jinling Hospital, which brings a powerful visual encoder. In the second stage, we integrate a textual encoder and conduct contrastive learning using 11,106 paired cardiac CT and radiology reports from Jinling Hospital. Specifically, to facilitate the effective contrastive training, free-text reports are first standardized into structured sentences that reflect the significant pathological findings, which are utilized to construct case-specific pathology vectors summarizing diagnostic attributes. We then compute the cosine similarity between pathology vectors of different cases to represent their semantic affinity, which is leveraged to build a soft supervision matrix for contrastive training instead of the traditional one-hot matrix. This strategy enables Cardiac-CLIP to learn fine-grained visual–language relationships corresponding to cardiac pathological semantics \cite{zhang2021self, wang2021knowledge, hu2021hierarchical}.

To comprehensively evaluate the effectiveness of Cardiac-CLIP, we perform extensive experiments on cardiac imaging data from diverse sources, utilizing a total of 32,118 CT scans. On the one hand, we incorporate publicly available chest CT datasets such as NLST \cite{7_NLSTdata_2011} and CT-RATE \cite{5_5_Hamamci2024} to ensure reproducibility. On the other hand, to validate the clinical applicability of Cardiac-CLIP, we perform multi-center evaluations on 6,722 real-world cardiac CT data collected from 12 different hospitals. We organize our evaluation into three major aspects: cardiovascular abnormality classification, information retrieval, and clinical analysis. Specifically, the clinical analysis tasks include prospective prediction of acute coronary syndrome (ACS), functional coronary stenosis (FCS) assessment, and coronary artery calcium (CAC) grading, which are significantly important and notoriously difficult to diagnose in real-world practice. Experimental results demonstrate that Cardiac-CLIP not only achieves state-of-the-art performance on public datasets, outperforming existing 3D medical foundation models by a large margin, but also exhibits remarkable capacity in complex real-world clinical analysis, reflecting its strong clinical utility. Furthermore, Cardiac-CLIP exhibits emergent capabilities, demonstrating outstanding performance in the diagnosis and prognosis of diseases like ACS that are not explicitly included during pre-training, thereby highlighting its strong generalization potential for previously unseen cardiovascular diseases.

\section{Data}
As the first attempt to develop a foundation model specifically for cardiac CT images, we construct a large-scale dataset comprising chest CT, CACCT and CCTA scans to enable robust pre-training and comprehensive clinical evaluation. The whole dataset is built upon publicly available data as well as multi-center real-world clinical data collected from 12 hospitals. The details of our dataset are shown in Fig.~\ref{fig:data}. In the following parts, we introduce the datasets utilized in our study.

\subsection{Publicly Available Data}
To facilitate the effective training and comprehensive evaluation of the foundation model, large-scale data is both critical and indispensable. However, due to the absence of publicly available large-scale cardiac CT dataset, we incorporate two chest CT datasets, NLST and CT-RATE, into our study. Both datasets are preprocessed using a pre-trained RetinaNet \cite{3_9_NC_Chao2021} to extract cardiac regions and remove irrelevant background structures. Specifically, the NLST dataset is mainly utilized to perform MAE pre-training during the first stage to enrich the training data and enhance the visual encoder, while CT-RATE is only utilized during the evaluation phase. By leveraging these publicly available datasets, we not only enrich the training data but also enhance the transparency and reproducibility of our experiments. The detailed descriptions are provided below. Notably, for the convenience of the following description, we denote NLST and CT-RATE as \textbf{O}(pen-source)\textbf{D}(ata)-1 and OD-2, respectively.

\textbf{OD-1: NLST Dataset} \\
The National Lung Screening Trial (NLST) is a landmark randomized multi-center study conducted under the sponsorship of the U.S. National Institutes of Health (NIH). In this work, we utilize low-dose computed tomography (LDCT) scans from the NLST dataset, obtained from The Cancer Imaging Archive (TCIA). To focus on cardiac anatomy, all images are preprocessed using a cardiac region cropping procedure \cite{3_9_NC_Chao2021}, leading to 116,313 CT images from 26,000 subjects. For the MAE pre-training stage, we utilize 114,248 cardiac-cropped images from 23,935 subjects, excluding those designated as the test data in downstream cardiovascular abnormality classification experiments. Specifically, for the classification evaluation, we also follow the data selection and partitioning strategy outlined in \cite{3_9_NC_Chao2021}, leading to 29,109 images for fine-tuning, 1,008 for validation, and 2,065 for testing.

\vspace{3mm}
\textbf{OD-2: CT-RATE Dataset} \\
The CT-RATE dataset \cite{5_5_Hamamci2024} comprises 25,692 chest CT scans from 21,304 patients, expanded to 50,188 images through various reconstruction protocols. In our work, we only select images reconstructed using Protocol 1 and crop the cardiac region, resulting in 23,331 cardiac-cropped images. As depicted in Fig.~\ref{fig:data}\textbf{(a)} and Fig.~\ref{fig:data}\textbf{(b)}, the CT-RATE is only utilized in the downstream evaluation for cardiovascular abnormality classification and information retrieval, with no samples included in the two-stages pre-training. Specifically, in accordance with the original dataset split strategy \cite{5_5_Hamamci2024}, 1,463 images are employed for final evaluation, while 90\% of the remaining 21,868 images are used for fine-tuning and 10\% for validation. As shown in Fig.~\ref{fig:data}\textbf{(c)}, this dataset encompasses four cardiovascular abnormalities commonly identifiable in chest CT: coronary artery calcification, aortic calcification, cardiomegaly, and pericardial effusion.

\subsection{Real-world Clinical Data}
To develop a clinically applicable foundation model, we collect cardiac CT data from 12 different hospitals for training and evaluation. Due to privacy constraints, these datasets are not publicly available. For clarity and convenience in subsequent descriptions, we utilize the \textbf{P}(rivate) \textbf{D}(ata)-\emph{ID} to denote these datasets. The information of these datasets is summarized below.

\textbf{PD-1: Jinling Hospital Dataset}
This dataset is collected from Jinling Hospital, Affiliated Hospital of Medical School, Nanjing University, which consists of two temporally distinct cohorts. One of the cohorts includes Chest CT, CACCT and CCTA collected from 5,818 patients between year 2007 to 2020. This cohort is utilized in both the two-stage pre-training and the internal evaluation. Specifically, 16,641 images (5,535 chest CT and 11,106 cardiac CT) from 4,708 patients are used for MAE pre-training, among which 11,106 cardiac CT images from 3,294 patients are further involved in the contrastive pre-training. Meanwhile, 1,663 images (798 CCTA and 865 CACCT) from 946 patients are utilized as contemporaneous internal evaluation data. Notably, as shown in Fig.~\ref{fig:data}\textbf{(c)}, this cohort contains 7 different cardiovascular conditions.

To evaluate the robustness of model on temporally shifted data, we involve a dedicated cohort that contains CCTA images from 2,576 patients collected in 2024 from the same hospital (Jinling Hospital). This cohort contains 5 different cardiovascular conditions that are also present in the prior cohort. Notably, aortic calcification and atherosclerosis are excluded due to changes in clinical reporting practices. This temporally-shifted cohort provides a benchmark for evaluating model stability over time, considering that patient characteristics and clinical workflows may evolve subtly.

\begin{figure}[t]
\centering
    \includegraphics[width=1\linewidth]{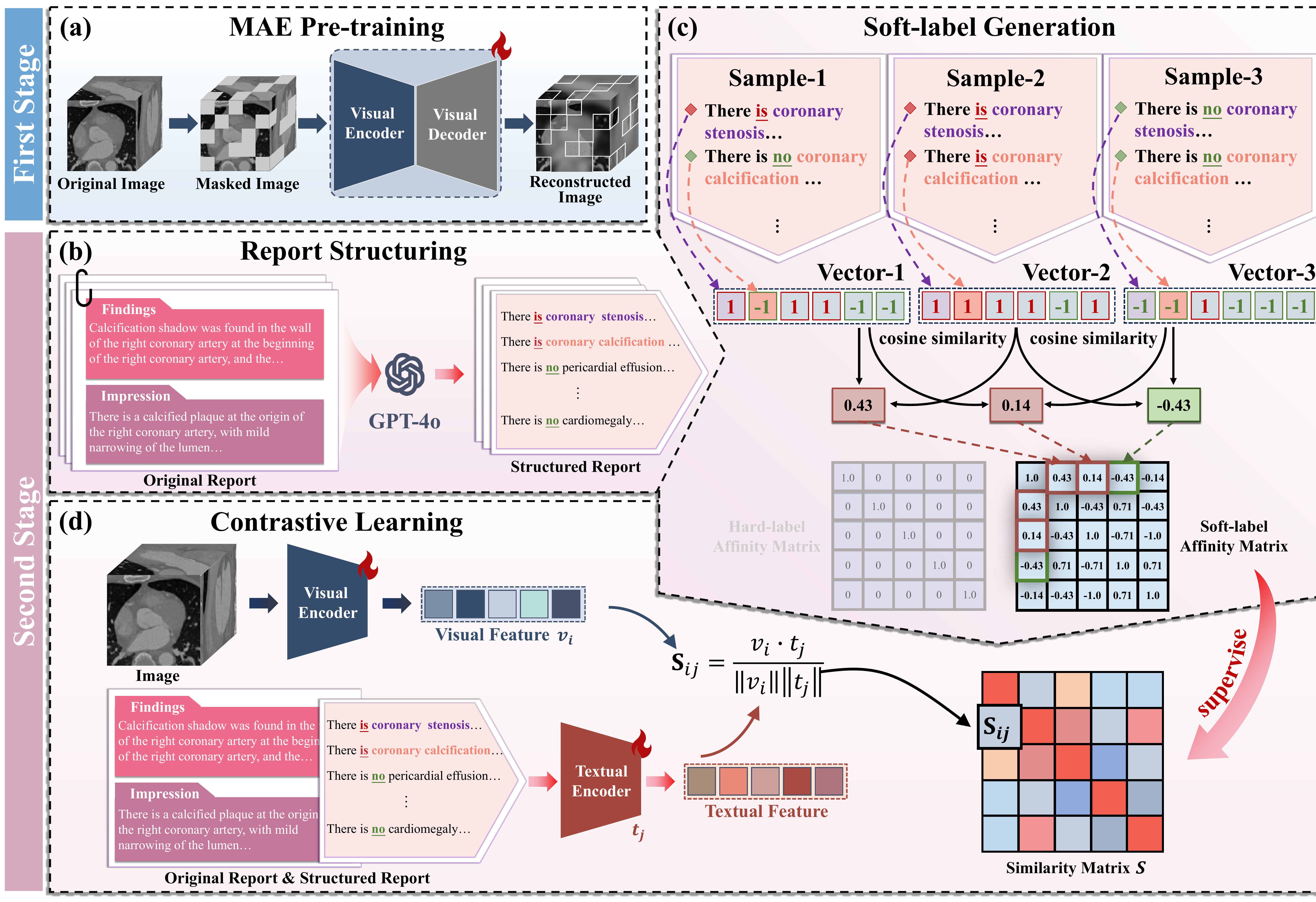}
\caption{The whole training pipeline of our Cardiac-CLIP. \textbf{(a)} The first stage pre-training is performed based on MAE framework, which brings a powerful visual encoder. \textbf{(b)} The process of report standardization. \textbf{(c)} Soft-label generation process. \textbf{(d)} The second stage pre-training via multi-modal contrastive learning. }
\label{fig:framework}
\end{figure}

\textbf{PD-2: Jiangsu Province Hospital Dataset} \\
This retrospective dataset is created from the hospital database for a study on diabetes mellitus complicated by acute coronary syndrome (ACS), comprising 650 diabetic patients with paired CCTA scans and diagnostic reports. The dataset includes a study group of 261 patients who develop ACS symptoms during a two-year follow-up and a control group of 389 patients. Meanwhile, as illustrated in Fig.~\ref{fig:data}\textbf{(c)}, 7 abnormal cardiovascular findings and conditions are involved in this dataset, \emph{e.g.,} coronary stenosis, coronary calcification, aortic calcification, atherosclerosis, cardiomegaly, pericardial effusion, and coronary artery disease. This dataset could support various downstream evaluations, including classification, information retrieval, and prospective risk prediction of acute coronary syndrome.

\textbf{PD-3: Multi-center CAC Dataset} \\
The Multi-center CAC Dataset comprises 1,311 non-contrast CT scans acquired from three clinical institutions, including 862 CACCT scans from Nanjing First Hospital, 260 CACCT from Northern Jiangsu People's Hospital, and 189 chest CT scans from the Affiliated Hospital of Jiangnan University. Cardiac region cropping is applied to all chest CT scans. Each scan is annotated with a coronary artery calcium score based on the standardized Agatston scoring protocol, according to which the calcification grade is determined. This dataset provides a benchmark for assessing the capacity of foundation models on coronary artery calcium grading.

\textbf{PD-4: Multi-center FCS Dataset} \\
The Multi-center FCS Dataset comprises 522 high-quality CCTA scans from 391 patients, collected across 9 different clinical institutions. As shown in Fig.~\ref{fig:data}(a), according to the index in the figure, these institutions are: (1) Sir Run Run Shaw Hospital, School of Medicine, Zhejiang University; (2) Beijing Anzhen Hospital, Capital Medical University; (3) Shanghai Sixth People's Hospital; (4) Jinling Hospital, Affiliated Hospital of Medical School, Nanjing University; (5) Guangdong Provincial People's Hospital; (6) Xijing Hospital, Air Force Medical University; (7) Peking Union Medical College Hospital; (8) Nanjing First Hospital; (9) The Affiliated Taizhou People's Hospital of Nanjing Medical University. Each case in the hospital includes corresponding invasive fractional flow reserve (FFR) scores, which indicate the presence of functional coronary stenosis. Within this dataset, 220 cases are confirmed to have functional stenosis. Generally speaking, this dataset is employed to assess the ability of each foundation model to directly diagnose functional coronary stenosis from CCTA images.

\section{Result}
In this section, we present the experimental evaluation of the proposed Cardiac-CLIP. Our goal is to develop a foundation model for cardiac CT that is applicable to real-world clinical scenarios. To this end, the ability to directly handle 3D volumetric images is critical and indispensable. Accordingly, we perform experiments exclusively on 3D medical foundation models, as 2D models are inherently incapable of processing volumetric clinical imaging data. Specifically, we compare Cardiac-CLIP against three representative 3D foundation models, Merlin \cite{5_4_blankemeier2024merlin}, CT-CLIP \cite{5_5_Hamamci2024} and M3DNet \cite{5_6_bai2024m3d} which have obtained substantial attention for their prior successes in medical vision tasks. Meanwhile, for the fine-tuning evaluation, we also include the traditional deep learning model, 3D-ViT, as the baseline methods for comparison. Our experiments are divided into three aspects: cardiovascular abnormality classification, information retrieval, and clinical analysis. Generally speaking, Cardiac-CLIP consistently outperforms prior foundation models, demonstrating superior performance across all evaluation tasks.

\subsection{Problem formulation}
In our experiments, we primarily evaluate the performance of each model in both zero-shot and fine-tuning settings. Here, we first present the pipeline of these two settings utilized in the cardiovascular abnormality classification (Sec.~\ref{sec:ex_cls}) and clinical analysis (Sec.~\ref{sec:ex_ana}). The detailed pipeline of the information retrieval experiments is described separately in Sec.~\ref{sec:ex_ret}.

\subsubsection{Zero-shot Evaluation}
\begin{figure}[htpb]
\begin{center}
    \includegraphics[width=1\linewidth]{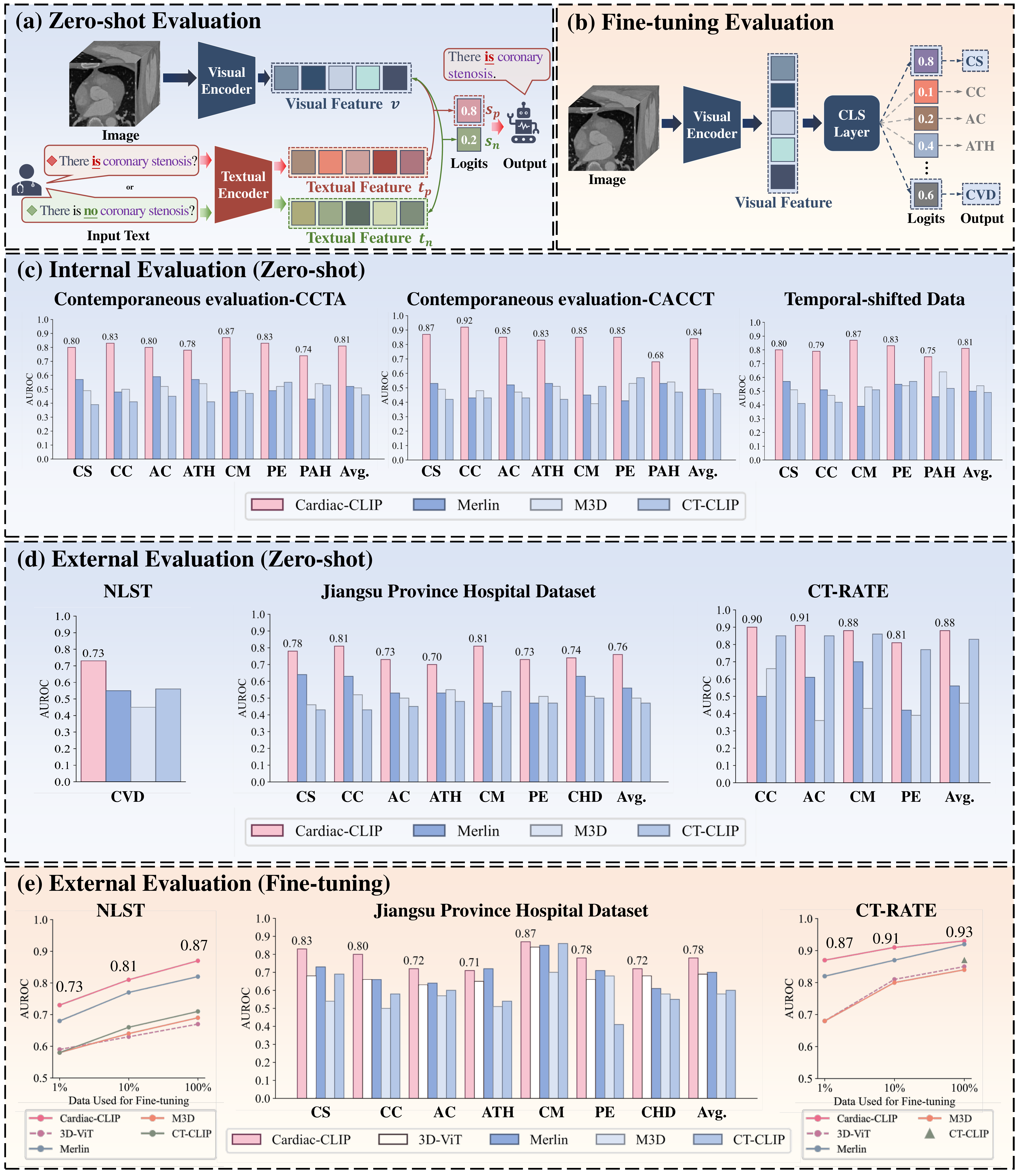}
    \end{center}
    \vspace{-3mm}
    \caption{The experiments for cardiovascular abnormality classification. \textbf{(a)}-\textbf{(b)}, the pipeline of zero-shot evaluation and fine-tuning evaluation, respectively. \textbf{(c)} The experimental results for internal evaluation. \textbf{(d)}-\textbf{(e)} The experimental results for external evaluation in zero-shot and fine-tuning manner, respectively. The abbreviations in the figure are defined as follows: \textbf{CS}: Coronary Stenosis, \textbf{CC}: Coronary Calcification, \textbf{AC}: Aortic Calcification, \textbf{ATH}: Atherosclerosis, \textbf{CM}: Cardiomegaly, \textbf{PE}: Pericardial Effusion, \textbf{PAH}: Pulmonary Arterial Hypertension, \textbf{CHD}: Coronary Heart Disease, \textbf{CVD}: Cardiovascular Disease. Meanwhile, \textbf{Avg.} denotes the average AUROC score across the corresponding cardiovascular abnormalities. Notably, in the bottom-right plot, the green triangle indicates the accuracy of CT-CLIP fine-tuned on the full dataset, which is directly extracted from its original paper \cite{5_5_Hamamci2024}.}
    \label{fig:class}
\end{figure}

As shown in Fig.~\ref{fig:class}\textbf{(a)}, in the zero-shot evaluation, we directly utilize the pre-trained parameters for inference without any additional fine-tuning. Taking ``coronary stenosis'' as an example, we construct two textual prompts: ``There is coronary stenosis'' and ``There is no coronary stenosis'', which are encoded by the textual encoder of the foundation model to obtain the corresponding textual features, denoted as $t_p$ and $t_n$, respectively. Simultaneously, the associated image is processed by the visual encoder to extract its visual representation $v$. Next, cosine similarities are calculated between the visual feature $v$ and each of the textual features, resulting in similarity scores $s_p = \cos(v, t_p)$ and $s_n = \cos(v, t_n)$. Then, the final prediction is determined by comparing $s_p$ and $s_n$. The prompt with higher similarity score will be selected as the final prediction, indicating whether the model identifies the presence or absence of the coronary stenosis in the image.

\subsubsection{Fine-tuning Evaluation}
As shown in Fig.~\ref{fig:class}\textbf{(b)}, in the fine-tuning evaluation, we extract the visual encoder from the pre-trained foundation model and append a classification head to form a complete classification network. The entire network is then fine-tuned on the target dataset, and its performance is assessed accordingly. In this way, we evaluate whether the pre-trained foundation model provides a strong initialization and transferable visual encoder for downstream tasks.

\subsection{Cardiovascular Abnormality Classification}
\label{sec:ex_cls}
To evaluate the performance of Cardiac-CLIP in cardiovascular abnormality classification, we conduct experiments using both internal and external datasets to assess its effectiveness and robustness.

\subsubsection{Internal evaluation}
We first perform the internal evaluation of Cardiac-CLIP using data from Jinling Hospital. Specifically, in addition to assess classification accuracy on samples collected contemporaneously with the training data, we also evaluate the performance on temporally shifted samples to validate its robustness against distributional shifts over time.

\textbf{Evaluation on Contemporaneous Data} \\
We evaluate our model on contemporaneous evaluation cohort from Jinling Hospital dataset. As shown in the first two histograms in Fig.~\ref{fig:class}\textbf{(c)}, Cardiac-CLIP consistently outperforms previous methods on 7 different cardiovascular abnormalities. Notably, although M3DNet and CT-CLIP are regarded as general-purpose and comprehensive 3D medical foundation models, they fail to deliver satisfactory performance on cardiac CT, with AUC scores similar to that of random guessing. These results underscore the limitations of existing 3D medical foundation models and highlight the necessity of the proposed Cardiac-CLIP.

\textbf{Evaluation on temporally shifted Data} \\
To assess the robustness of Cardiac-CLIP to temporal distribution shifts, we conduct the evaluation in a zero-shot setting on temporal-shifted cohort of Jinling Hospital dataset, which consists of CCTA scans collected from the same institution but during a different time period compared to the training data. As shown in the third histogram in the Fig.~\ref{fig:class}\textbf{(c)}, Cardiac-CLIP maintains strong performance across five common cardiovascular conditions, with only marginal degradation compared to results on contemporaneous data, demonstrating the model’s stability over time and its potential for sustained clinical applicability. The experimental results highlight the robustness of Cardiac-CLIP in retaining diagnostic reliability under temporal shift, which is critical for clinical deployment in real-world scenarios.

\subsubsection{External evaluation}
We evaluate the performance of our Cardiac-CLIP on three external dataset: NLST, Jiangsu Province Hospital Dataset and CT-RATE. Although the NLST dataset is included in the first-stage pre-training, only the image data is involved, without any semantic annotations. Therefore, NLST is treated as the external data for the cardiovascular abnormality classification task. Specifically, we perform experiments in both zero-shot and fine-tuning settings, which not only evaluates the generalization capacity but also verifies the transferability of each foundation model.

\textbf{Zero-shot Evaluation}
\label{sec:ex_cls_ze}
As illustrated in Fig.~\ref{fig:class}\textbf{(d)}, Cardiac-CLIP consistently demonstrates strong generalization performance when evaluated in a zero-shot setting on three external datasets, significantly outperforming all compared methods. On the Jiangsu Province Hospital dataset, Cardiac-CLIP achieves notably higher AUROC scores, with increases exceeding 10\% across all involved abnormalities. In particular, for cardiomegaly and pericardial effusion, Cardiac-CLIP surpasses the second-best model by more than 30\%. Notably, despite the CT-RATE dataset not being included in either stage of pre-training, Cardiac-CLIP achieves the highest AUROC scores on this dataset, even outperforming CT-CLIP that is trained directly on the CT-RATE training set. These findings highlight the strong generalization capability of Cardiac-CLIP, underscoring its advantages over existing 3D foundation models in cardiovascular abnormality classification tasks.

\textbf{Fine-tuning Evaluation}
To further evaluate the effectiveness and transferability of Cardiac-CLIP, we fine-tune all the models on the external datasets and evaluate their performance. As shown in Fig.~\ref{fig:class}\textbf{(d)}, Cardiac-CLIP consistently achieves the highest AUROC scores across all datasets.

Specifically, on the two large-scale datasets, NLST and CT-RATE, we conduct controlled experiments by varying the proportion of labeled training data used for fine-tuning, and report the average AUROC across all abnormalities. Remarkably, even when fine-tuned with only 10\% of the training data, Cardiac-CLIP achieves satisfactory performance, highlighting its superior data efficiency. For the Jiangsu Province Hospital dataset, which is relatively small in scale, we perform fine-tuning using the full training set. Cardiac-CLIP still demonstrates obvious advantages over compared methods. These experimental results confirm that pre-training on large-scale data enables Cardiac-CLIP to learn transferable and clinically meaningful representations, providing a robust initialization that facilitates downstream adaptation with minimal supervision.

\begin{figure}[htpb]
\begin{center}
    \includegraphics[width=1\linewidth]{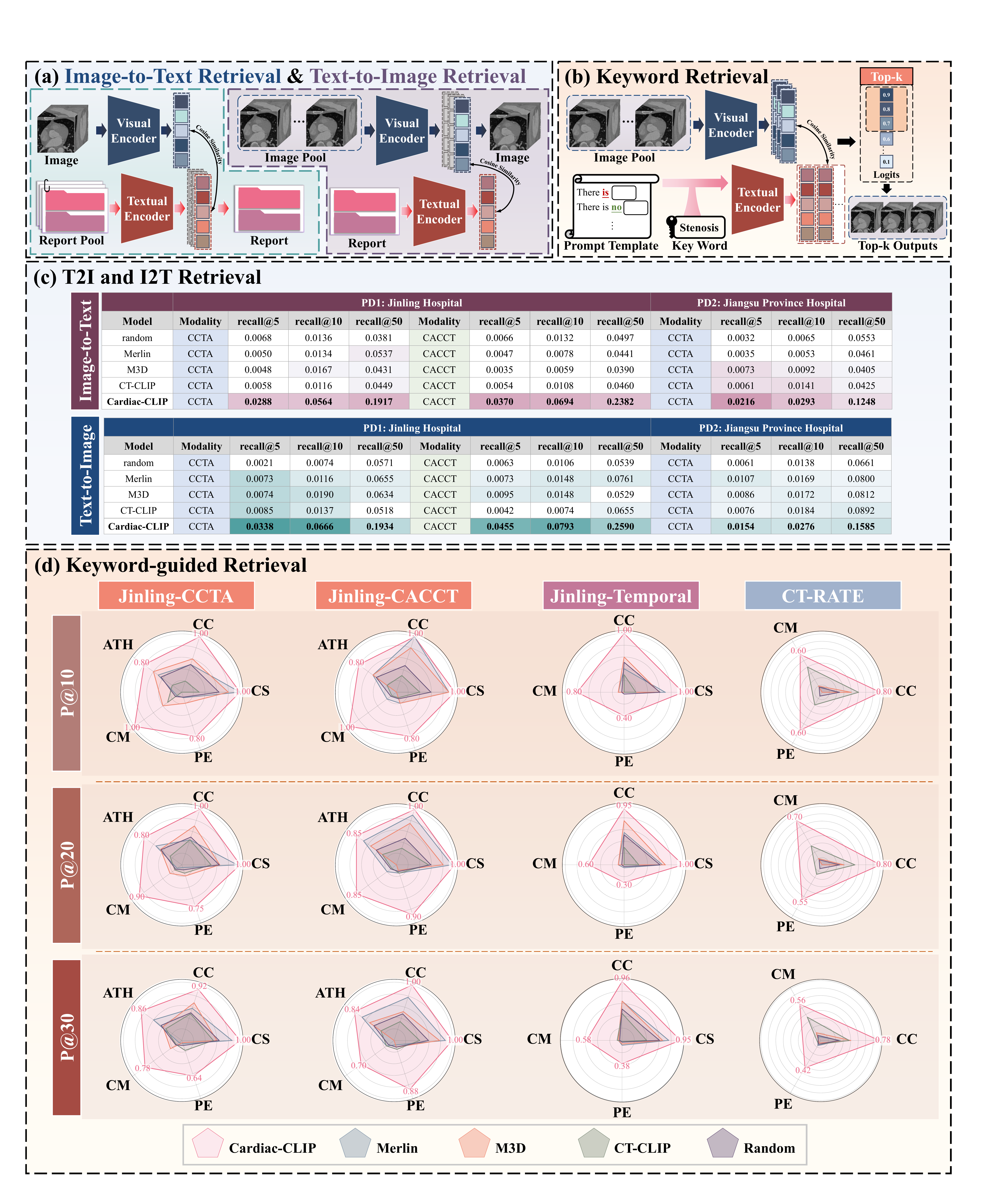}
    \end{center}
    \caption{The experiments for information retrieval. \textbf{(a)}-\textbf{(b)}, the pipeline of image-to-text retrieval, text-to-image retrieval and keyword-guided retrieval, respectively. \textbf{(c)} The experimental results for text-to-image and image-to-text retrieval. \textbf{(d)} The experimental results for keyword-guided retrieval, in which \textbf{CS} denotes coronary stenosis, \textbf{CC} denotes coronary calcification, \textbf{ATH} denotes Atherosclerosis, \textbf{CM} denotes Cardiomegaly and PE denotes Pericardial Effusion.}
    \label{fig:retrieval}
\end{figure}

\subsection{Information Retrieval}
\label{sec:ex_ret}
Information retrieval plays a pivotal role in supporting clinical decisions by enabling efficient access to relevant diagnostic content, which reflects the capacity of foundation models for cross-modal alignment. Here, we conduct experiments to evaluate Cardiac-CLIP on three different retrieval tasks: image-to-text retrieval, text-to-image retrieval, and keyword-guided retrieval. The differences among these experimental settings are illustrated in Fig.~\ref{fig:retrieval}\textbf{(a)} and Fig.~\ref{fig:retrieval}\textbf{(b)}. Specifically, the image-to-text retrieval task evaluates whether the model can accurately retrieve the most relevant textual report given an input 3D cardiac image, thereby assessing its ability to align visual features with semantic content. Conversely, text-to-image retrieval utilizes the radiology report to query the most relevant cardiac image from a predefined image pool, highlighting the capacity to ground textual semantics in visual representations. For these two settings, we employ Recall@K (R@5, R@10, and R@50) as evaluation metrics. Finally, keyword-guided retrieval provides a more clinically realistic setting, in which a single diagnostic phrase ( \emph{e.g.,} ``There is coronary artery calcium'') is used as a query to retrieve the image with corresponding cardiovascular abnormality, thereby evaluating the capacity to capture disease-specific patterns. For this setting, we adopt Precision@K (P@5, P@10, and P@50) to evaluate retrieval accuracy, which measures the proportion of correctly matched samples among the top-K retrieved candidates. Notably, all experiments in this section are conducted using the original parameters of each foundation model in the zero-shot setting, reflecting their capacity for cross-modal alignment.

\subsubsection{Image-to-Text Retrieval}
In the image-to-text retrieval experiments, model needs to select the correct radiology report corresponding to a given cardiac image from a predefined report pool. The retrieval pool comprises the entire test set, containing 1,663 structured radiology reports from the Jinling Hospital dataset (contemporaneous cohort) and 650 reports from the Jiangsu Province Hospital dataset for two experiments, respectively. As shown in the first part of Fig.~\ref{fig:retrieval}\textbf{(c)}, Cardiac-CLIP achieves the highest performance across both datasets, substantially outperforming all prior methods. These results highlight the superior capability of our method in aligning visual features with diagnostic textual descriptions.

\subsubsection{Text-to-Image Retrieval}
In the text-to-image retrieval experiment, model is required to retrieve the correct images corresponding to the given radiology report from a predefined image pool. The test set is constructed following the same protocol used in the image-to-text retrieval experiments. As shown in the second part of Fig.~\ref{fig:retrieval}\textbf{(c)}, the experimental results show that Cardiac-CLIP consistently achieves the best performance across all R@5, R@10, and R@50 metrics, further validating its capability in grounding textual semantics within visual representations.

\subsubsection{Keyword-guided Retrieval}
In the keyword-guided retrieval task, model will select the relevant cardiac CT according to the query phrase from a predefined image pool. Specifically, we select 5 representative cardiovascular findings and conditions \emph{e.g.,} coronary stenosis, coronary calcification, atherosclerosis, cardiomegaly and pericardial effusion as the keywords to construct the query phrases. Experiments are conducted on both the contemporaneous and temporally shifted cohorts of the Jinling Hospital dataset, as well as the CT-RATE dataset. As shown in the radar plot in Fig.~\ref{fig:retrieval}\textbf{(d)}, Cardiac-CLIP consistently achieves the highest retrieval performance across the included 5 abnormalities. Notably, consistent with the findings in Sec.~\ref{sec:ex_cls_ze}, Cardiac-CLIP outperforms the CT-RATE-trained foundation model (CT-CLIP) on the CT-RATE dataset. Considering that CT-RATE is completely not included in our pre-training phase, this result further underscores the strong zero-shot generalization capability of Cardiac-CLIP. Generally speaking, these information retrieval experiments reflect the superior capacity of our Cardiac-CLIP in cross-modal alignment.

\begin{figure*}[htpb]
\begin{center}
    \includegraphics[width=1\linewidth]{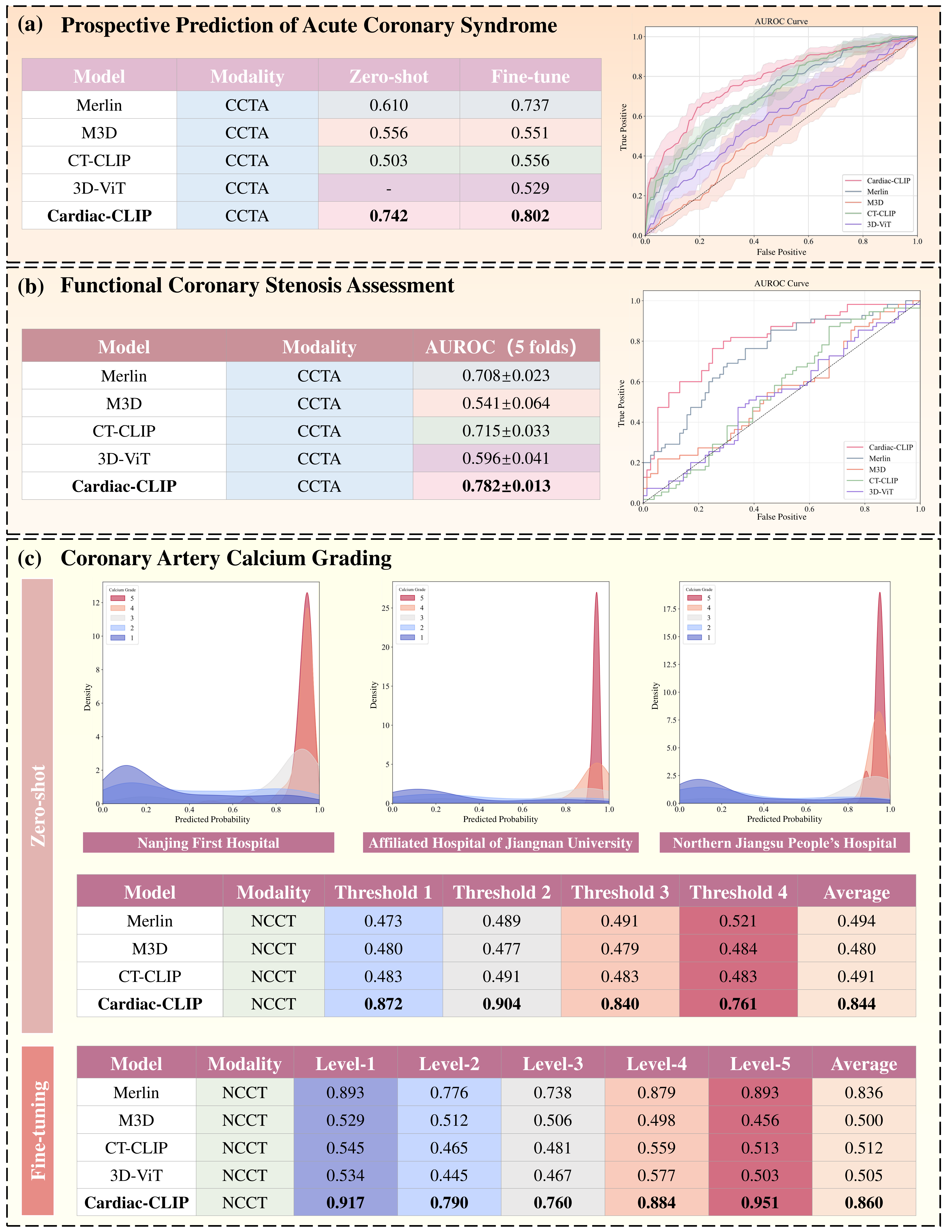}
    \end{center}
    \caption{The experimental results for clinical analysis. According to these experiments, our Cardiac-CLIP demonstrates remarkable capacity and broader potential in supporting clinical decision-making }
    \label{fig:analysis}
\end{figure*}

\subsection{Clinical Analysis}
\label{sec:ex_ana}
To assess the applicability of Cardiac-CLIP in real-world clinical scenarios, we conduct a series of in-depth clinical analyses beyond the standard diagnostic classification. Specifically, we perform experiments in three challenging clinical tasks: prospective risk prediction for acute coronary syndrome, diagnosis of functional coronary stenosis, and coronary artery calcium grading. These tasks are known to be difficult due to subtle imaging features, complex physiological implications, and the scarcity of explicit annotations in routine radiology reports. By demonstrating strong performance in these tasks, Cardiac-CLIP illustrates its capacity to support clinical decision-making and highlights the broader potential for cardiovascular diagnosis.

\subsubsection{Prospective Risk Prediction for Acute Coronary Syndrome}
\label{sec:ana_acs}
Acute coronary syndrome (ACS) is a critical condition that demands timely and accurate prospective risk prediction \cite{bergmark2022acute, al2020machine}. However, its early-stage imaging manifestations are often subtle, making prospective prediction highly challenging. This experiment is conducted on Jiangsu Province Hospital dataset. Each model is required to predict whether ACS will occur within the subsequent two years for 650 patients, based on their prior CCTA scans.

As shown in Fig.~\ref{fig:analysis}\textbf{(a)}, Cardiac-CLIP achieves remarkable performance in both zero-shot and fine-tuning settings. Specifically, the traditional deep learning baseline, 3D-ViT, achieves only a limited AUROC of 0.529 under fully supervised training. Similarly, M3DNet and CT-CLIP only obtain AUROC scores of approximately 0.55 after fine-tuning, which are close to random guessing. These results further underscore their inadequacy in ACS assessment. In contrast, Cardiac-CLIP outperforms these fine-tuned models even in the zero-shot settings. Moreover, after the fine-tuning, the accuracy of Cardiac-CLIP is further improved to 0.802, reflecting its great potential in prospective risk prediction for acute coronary syndrome.

\subsubsection{Diagnosis of Functional Coronary Stenosis}
\label{sec:ana_fcs}
Functional coronary stenosis refers to the physiologically significant narrowing of the coronary arteries that leads to impaired myocardial perfusion, which holds substantial clinical importance in guiding treatment decisions. Existing diagnostic methods rely on invasive fractional flow reserve (FFR) measurements or noninvasive CT-derived FFR (CT-FFR) techniques \cite{ko2017noninvasive}, which are often constrained by complex workflows and sensitivity to segmentation errors. Here, to assess the ability of foundation models in diagnosing functional coronary stenosis, we conduct experiments using the Multi-center FCS Dataset in the fine-tuning setting with five-fold cross-validation. As shown in Fig.~\ref{fig:analysis}\textbf{(b)}, our Cardiac-CLIP achieves the best performance, with an average AUC of 0.782$\pm$0.013, significantly outperforming previous foundation models. Specifically, M3D only obtains AUROC score of 0.596$\pm$0.041, which is barely above random guessing. These results suggest that previous models lack sufficient cardiovascular domain knowledge, leading to suboptimal performance even after task-specific fine-tuning.

\subsubsection{Coronary Artery Calcium Grading}
\label{sec:ana_cac}
Coronary artery calcium serves as a key marker for atherosclerotic burden and cardiovascular risk stratification. Conventional CAC grading relies on specialized software. In this task, we evaluate the ability of Cardiac-CLIP to infer CAC grade in both zero-shot and fine-tuning settings.

Since explicit CAC scores are not consistently included in every radiology report and it is difficult to directly align continuous values with visual features, we exclude the information of CAC score when constructing the textual training data. As a result, explicit CAC grades are not available during the pre-training phase of Cardiac-CLIP, making zero-shot CAC grading particularly difficult. To address this issue, we adopt a proxy-based approach that leverages predicted probability from foundation model to assess calcification severity, which measures the inherent implicit knowledge stored in the model \cite{hinton2015distilling, xu2020knowledge}. Specifically, for each CT scan, we compute the cosine similarity between its visual embedding and the textual prompt ``There is Coronary Artery Calcium''. The similarity score reflects the model’s confidence regarding the presence of coronary calcification within the input CT. As shown in the probability distribution map within Fig.~\ref{fig:analysis}\textbf{(c)}, we find that the confidence scores assigned by Cardiac-CLIP correlate well with calcium grade. Specifically, cases with the highest level of calcification (Grade-5) exhibit confidence scores predominantly clustered near 1.0, while those with the lowest calcium level (Grade-1) are assigned scores concentrated near 0. This trend suggests that the model has effectively captured semantic knowledge related to coronary artery calcium, producing increasingly confident predictions as the degree of calcification becomes more severe.

According to these finds, we further investigate the capacity of different foundation models in CAC grading through the Ordinal AUROC metric, which extends the standard AUROC to ordered multi-class settings by computing AUROC scores across multiple binary classification thresholds. Specifically, as illustrated in the first table of Fig.~\ref{fig:analysis}\textbf{(c)}, we assess four thresholds that divide the five CAC levels into increasingly severe categories (\emph{e.g.}, Grade 1 vs. Grade 2–5, Grade 1–2 vs. Grade 3–5, \emph{etc}). The results show that Cardiac-CLIP achieves consistently high scores across all thresholds, demonstrating its capacity to distinguish between different levels of calcification. This suggests that the model encodes rich domain knowledge relevant to cardiovascular images. In contrast, existing foundation models exhibit near-random performance under the same zero-shot setting, reflecting a lack of cardiovascular-specific representation ability.

Meanwhile, we also fine-tune each foundation model to evaluate their performance under task-specific supervision. In this condition, we attach a five class classifier after the visual encoder to directly predict the five different calcium grades. As shown in the second table in Fig.~\ref{fig:analysis}\textbf{(c)}, Cardiac-CLIP achieves the highest AUROC scores, outperforming prior methods by a substantial margin. Notably, even after fine-tuning, M3DNet, CT-CLIP, and 3D-ViT continue to exhibit suboptimal performance, with AUROC scores close to random guessing, which further highlights their limited capacity to capture relevant domain knowledge.

\subsection{Ablation Study}
\begin{figure*}[t]
\begin{center}
    \includegraphics[width=1\linewidth]{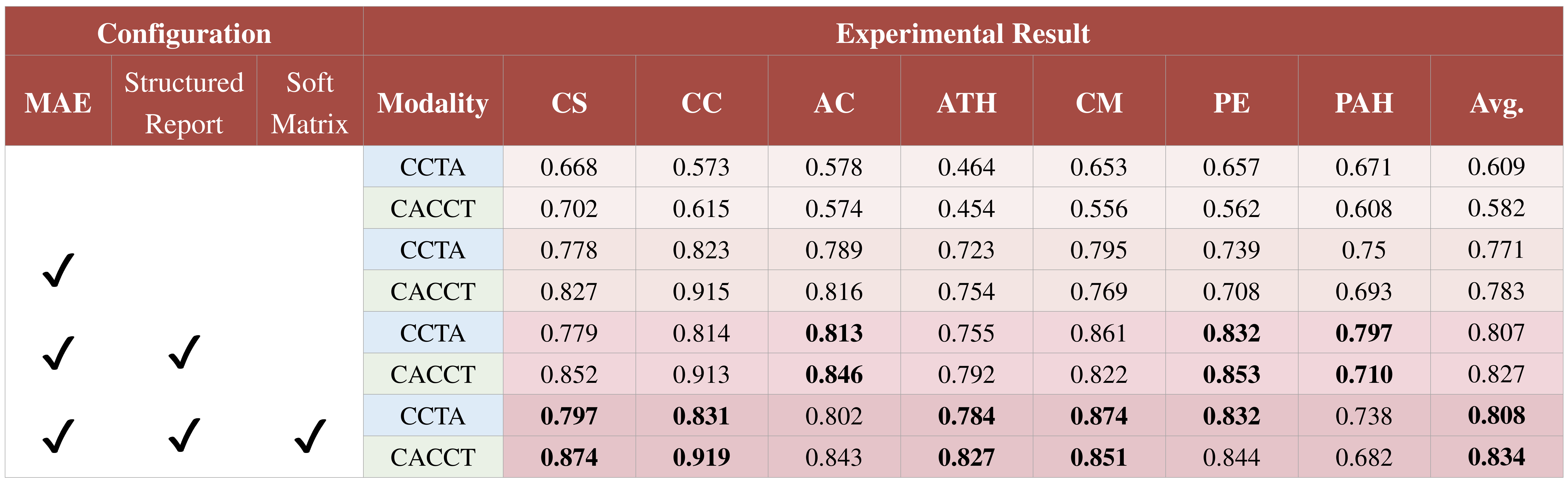}
    \end{center}
    \caption{The experimental results for ablation study. The abnormality abbreviations are similar to the previous part. \textbf{CS} denotes coronary stenosis, \textbf{CC} denotes Coronary Calcification, \textbf{AC} denotes Aortic Calcification, \textbf{ATH} denotes Atherosclerosis, \textbf{CM} denotes Cardiomegaly, \textbf{PE} denotes Pericardial Effusion and \textbf{PAH} denotes Pulmonary Arterial Hypertension. Meanwhile, \textbf{Avg.} indicates the average AUROC score across the corresponding abnormality.}
    \label{fig:ablation}
\end{figure*}

To further investigate the effectiveness of different components of our method, we conduct a comprehensive ablation study on the contemporaneous evaluation cohort from the Jinling Hospital dataset. The experimental results are listed in Fig.~\ref{fig:ablation}, in which we compare the performance improvements brought by MAE pre-training, the structured report and soft-label matrix, respectively. Our Cardiac-CLIP is established when all these three components are utilized. As shown, each involved component all brings significant accuracy increases. The best performance are mainly delivered by the completed Cardiac-CLIP model. 

\section{Discussion}
In recent years, foundation models have garnered significant attention in the field of computer vision due to their impressive scalability and generalization capabilities. While similar efforts have emerged in the medical domain, most existing medical foundation models are limited to processing 2D image slices, with only a few capable of handling full 3D medical imaging data. In our study, we develop Cardiac-CLIP and compare its performance against several representative 3D medical foundation models. In experiments such as cardiovascular abnormality classification and information retrieval, Cardiac-CLIP consistently outperforms other methods, suggesting its superior visual perception capabilities in cardiovascular-related tasks.

Since our work aims to establish a clinically applicable foundation model, we also conduct experiments to evaluate the effectiveness of Cardiac-CLIP on three clinically significant and highly challenging cardiovascular tasks: prospective risk prediction for acute coronary syndrome, the diagnosis of functional coronary stenosis and coronary artery calcium grading. These three tasks are extremely challenging, where traditional deep learning methods cannot produce satisfactory performance even under fully supervised training. 

Specifically, ACS is a life-threatening cardiovascular condition that requires timely and accurate risk stratification to inform clinical decision-making. However, prospectively predicting the occurrence of ACS remains a challenging task due to its complex pathophysiology and often subtle imaging manifestations in the early stages. As stated in Sec.~\ref{sec:ana_acs}, traditional deep learning approaches, even when trained in a fully supervised manner, struggle to achieve reliable performance in this task. In contrast, Cardiac-CLIP, empowered by large-scale multi-modal pre-training, demonstrates strong capability in ACS risk assessment even in the zero-shot setting, highlighting its potential to support early diagnosis and improve clinical outcomes.

Functional coronary stenosis refers to the physiologically significant narrowing of the coronary arteries that leads to impaired myocardial perfusion. Unlike anatomical stenosis, which focuses solely on luminal narrowing, functional stenosis reflects the actual hemodynamic impact of the lesion and thus plays a more critical role in guiding revascularization. Therefore, accurate identification of functional coronary stenosis is critical for optimizing treatment strategies and avoiding unnecessary interventions. Existing diagnostic methods rely on the invasive FFR measurement or noninvasive CT-FFR technique \cite{ko2017noninvasive}, which contains multiple steps and is complex in procedure. Specifically, CT-FFR requires the computational fluid dynamics (CFD) simulation with the well-delineated coronary lumen in CCTA images \cite{min2012diagnostic}. However, achieving precise coronary lumen segmentation in the presence of calcification or severe stenoses remains technically challenging and error-prone. These segmentation inaccuracies can propagate through the CFD modeling pipeline, leading to potential misestimation of FFR values and thus may require manual correction of the segmentation results. As demonstrated in Sec.~\ref{sec:ana_fcs}, Cardiac-CLIP, empowered by large-scale pre-training, offers a novel approach for rapidly identifying patients who are potentially affected by functional coronary stenosis. These patients can then be recommended for further confirmatory test using invasive FFR measurement or CT-FFR. In this way, Cardiac-CLIP could enhance the diagnostic efficiency and support more accurate clinical decision-making.

Coronary artery calcium is a critical biomarker of atherosclerotic burden and plays a pivotal role in cardiovascular risk stratification. Accurate identification of calcified plaques and quantification of CAC scores are essential for assessing the level of coronary calcification, which provides important prognostic insight into coronary artery disease and associated cardiovascular events. In clinical practice, CAC grading is commonly performed on NCCT scans using specialized software that applies the Agatston scoring method based on predefined intensity thresholds \cite{agatston1990quantification, hecht20172016}. However, in the era of deep learning, progress has been limited by the lack of large-scale annotated datasets with precise CAC grade labels, which hinders the development of supervised learning approaches. As shown in Sec.~\ref{sec:ana_cac}, M3DNet, CT-CLIP and 3D-ViT still delivers inferior performance close to random guessing even after supervised fine-tuning, which highlights the inherent difficulty of this task and suggests that it is hard to learn accurate CAC prediction without sufficient prior domain knowledge. In contrast, Cardiac-CLIP achieves remarkable CAC grading performance even under zero-shot settings, which underscores its rich domain knowledge and suggests that it captures semantic cues of calcification through large-scale pre-training.

More importantly, diagnosing acute coronary syndrome, functional coronary stenosis, and the severity of coronary artery calcium requires complex clinical reasoning and additional computational steps that go beyond the direct interpretation of raw CT images. Although these diagnoses are ultimately derived from CT data, they typically rely on post-processing tools or specialized software to extract intermediate physiological or anatomical indicators. As a result, such diagnoses are rarely documented explicitly in routine radiology reports. Consequently, Cardiac-CLIP is trained without task-specific annotations or explicit labels for these conditions. Nevertheless, our model demonstrates strong performance across all three tasks. Specifically, for ACS prediction and CAC grading, Cardiac-CLIP achieves competitive accuracy even in zero-shot settings, underscoring its ability to generalize and infer high-level cardiovascular concepts. This emergent capability highlights the capacity of Cardiac-CLIP to internalize rich domain knowledge through large-scale pre-training, enabling advanced clinical reasoning beyond its original training objectives. Although specific terms such as ``acute coronary syndrome'', ``functional stenosis'', and detailed calcification grades are not explicitly included in the training data, the radiology reports used in second-stage contrastive learning contain related terminology, such as ``coronary'', ``stenosis'', and ``calcification''. By leveraging these domain-relevant implicit cues and the prior knowledge embedded in the textual encoder, Cardiac-CLIP demonstrates the potential of vision-language foundation models to support complex cardiovascular diagnoses in real-world clinical scenarios.

\begin{figure*}[t]
\begin{center}
    \includegraphics[width=1\linewidth]{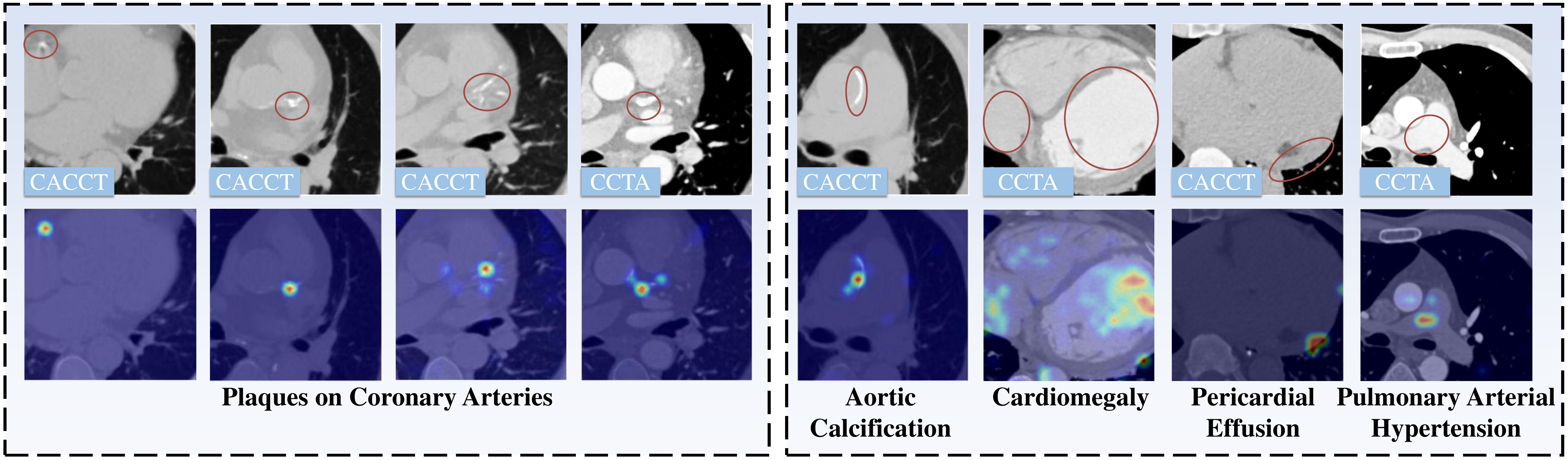}
    \end{center}
    \caption{Visualization results from the seventh layer of the visual encoder. Red circles indicate the regions associated with pathological findings. The text below each image represents the keyword used in the corresponding textual prompt. As illustrated by the heatmaps in the second row, Cardiac-CLIP effectively attends to clinically relevant areas, which is consistent with the actual lesion locations.}
    \label{fig:visualization}
\end{figure*}

To further assess the interpretability and effectiveness of Cardiac-CLIP, we visualize the feature maps from the seventh layer of the visual encoder. As the midpoint of the 12-layer backbone, this intermediate layer effectively balances spatial detail and high-level semantic abstraction, making it well-suited for visualization. Specifically, we use the prompt “There is [Keyword]” as textual input, where the utilized keywords are listed below each image. Given this text and the associated CT image, we extract the activation heatmaps from the seventh layer of the visual encoder using the visualization approaches proposed in \cite{chefer2021generic}. As shown in the first panel of Fig.~\ref{fig:visualization}, Cardiac-CLIP accurately highlights regions corresponding to calcified plaques. This suggests that, through large-scale pre-training, Cardiac-CLIP has acquired plaque-related knowledge, even in the absence of explicit localization information during the training process. The visualization results demonstrates the great potential of Cardiac-CLIP to assist clinical decision-making by providing reliable, human-interpretable visual cues in diagnostic scenarios.

\section{Method}

\subsection{Overall Architecture}
As illustrated in Fig.~\ref{fig:framework}, the overall architecture of Cardiac-CLIP consists of two main components, a visual encoder and a textual encoder. In the first stage, we only pre-train the visual encoder $E_v$ to learn robust image representations. In the second stage, we introduce the textual encoder $E_t$ and jointly perform multi-modal pre-training. The visual and textual features are utilized to compute cosine similarity scores, which capture the semantic relationships between images and corresponding textual descriptions.

\subsection{First-stage Pre-training}
As shown in Fig.~\ref{fig:framework}\textbf{(a)}, we pre-train the visual encoder of Cardiac-CLIP using the MAE framework in the first stage. This self-supervised strategy enables the model to learn semantically rich representations from volumetric cardiac image data without requiring manual annotations.

Following the standard MAE design, given a 3D input volume $I$, we divide it into $N$ non-overlapping patches. A random binary mask $M \in \left\{ {0,1} \right\}$ is then applied to the original volume, leading to a partially visible volume:
\begin{equation}
I_{mask} = I \cdot M.
\label{eq:mae_mask}
\end{equation}
The total number of patches satisfies $N = N_v + N_m$, where $N_v$ and $N_m$ denote the number of visible and masked patches, respectively. The visible patches $\mathbf{v}_{vis} \in \mathbb{R}^{B \times N_v \times C}$ are input to the visual encoder to produce latent representations, which are subsequently passed through a lightweight decoder to reconstruct the full volume, yielding $\mathbf{v}^{\text{recon}} \in \mathbb{R}^{B \times N \times C}$. Afterwards, to promote the representation capacity, the construction loss is computed using mean squared error (MSE) between the reconstructed and the original patches:
\begin{equation}
\mathcal{L}_{\text{mae}} = \left|| \mathbf{v}_{mask}^{\text{recon}} - \mathbf{g}_{mask}^{\text{recon}} \right||^2_2,
\label{eq:mae_loss}
\end{equation}
where $\mathbf{v}_{mask}^{\text{recon}}$ is the reconstructed masked patches and $\mathbf{g}_{mask}^{\text{recon}}$ indicates the corresponding ground-truth patches extracted from the original image $I$. By doing so, the visual encoder is encouraged to learn contextual representations that generalize well across downstream tasks, serving as a strong initialization for the subsequent stage to perform multi-modal alignment.

\subsection{Second-stage Pre-training}
In the second stage, we incorporate a textual encoder to enable multi-modal representation learning. To facilitate effective multi-modal pre-training, we first restructure the original free-text radiology reports into a unified, standardized format, yielding concise and interpretable textual inputs that are more appropriate to cross-modal alignment. Based on these structured reports, we construct a pathology vector for each case that summarizes key diagnostic attributes. These vectors are then used to compute a soft supervision matrix that captures inter-case semantic similarities, allowing the model to receive more fine-grained and informative supervision. Finally, contrastive learning is performed using image-text pairs to align visual and textual representations within a shared embedding space.

\subsubsection{Report Structuring}
Since the radiology reports are generated by different radiologists, they often exhibit considerable variability in format, resulting in the free-text documents and impeding efficient contrastive learning. To overcome this limitation, as illustrated in Fig.~\ref{fig:framework}\textbf{(b)}, we convert the original free-text reports into a unified, structured format, enabling more consistent and learnable textual representations for downstream multi-modal alignment. Specifically, we select a set of 7 representative cardiovascular abnormalities frequently mentioned in radiology reports (\emph{e.g.,} coronary stenosis, coronary calcification, aortic calcification, atherosclerosis, cardiomegaly, pericardial effusion, pulmonary arterial hypertension). Based on these abnormality categories, each original report is reformatted into a structured representation comprising 7 standardized textual statements. For each abnormality, if the report indicates its presence, we generate a positive prompt in the form of ``There is [abnormality-name]''. Conversely, if the report explicitly states the absence or makes no mention of the specific abnormality, we standardize the corresponding entry as ``There is no [abnormality-name].'' All these processes are conducted using the GPT-4o API \cite{hurst2024gpt}. By structuring the free-text report, we obtain semantically clearer and more uniform supervision signals, thereby facilitate the training of cross-modal alignment.

\subsubsection{Soft-label generation}
Based on the structured radiology reports, we construct a pathology vector for each case that summarizes the diagnostic attributes. Specifically, as depicted in Fig.~\ref{fig:framework}\textbf{(c)}, for each case, we construct a binary pathology vector $\mathbf{y} \in \left\{ {-1,1} \right\}^D$ over $D=7$ predefined cardiovascular attributes, where each element indicates the presence ($1$) or absence ($-1$) of a specific condition. Formally, the pathology vector for case $i$ is defined as:
\begin{equation}
\mathbf{y}_i^{(d)} =
\begin{cases}
1, & \text{if abnormality } d \text{ is reported as present in case } i \\
-1, & \text{otherwise}
\end{cases}
\label{eq:pathology_vector}
\end{equation}
Given a batch of $B$ samples, we compute pairwise cosine similarities between all pathology vectors to form a semantic affinity matrix $\mathbf{\tilde{A}} \in \mathbb{R}^{B \times B}$:
\begin{equation}
\mathbf{\tilde{A}}_{ij} = \frac{\mathbf{y}_i \cdot \mathbf{y}_j}{||\mathbf{y}_i|| \cdot ||\mathbf{y}_j||},
\label{eq:affinity_matrix}
\end{equation}
in which $y_i$ and $y_j$ denote the $i$-th and $j$-th sample in the batch. $\mathbf{\tilde{A}}$ reflects the semantic distances between different samples, where higher values denote greater similarity in diagnostic attributes. Unlike traditional contrastive learning frameworks that rely on binary hard labels, where sample pairs are assigned either 1 or 0, our soft-label matrix $\mathbf{\tilde{A}}$ provides a more expressive supervision signal that accounts for varying degrees of inter-case semantic similarity. By replacing the conventional binary label matrix with the soft-label affinity matrix, this approach enables the more fine-grained contrastive supervision, particularly benefiting the complex medical scenarios where inter-case relationships are not strictly dichotomous. Notably, during the contrastive learning process, both the original free-text radiology reports and their structured counterparts are utilized, with one randomly selected in each training iteration. This strategy not only provides more standardized and learnable textual data, but also preserves the full spectrum of diagnostic information inherent in the original reports.

\subsubsection{Contrastive Learning}
With the pre-trained visual encoder $E_v$ and initialized textual encoder $E_t$, we perform multi-modal contrastive learning to align image and textual representations in a shared embedding space. Given a batch of $B$ paired samples ${(x_i, e_i)}_{i=1}^B$, where $x_i$ denotes the input 3D cardiac volume and $e_i$ is the corresponding radiology report, the encoders generate visual features $v_i = E_v(x_i)$ and textual features $t_i = E_t(e_i)$.

As shown in Fig.~\ref{fig:framework}\textbf{(d)}, we first compute the cosine similarity between all pairs of visual and textual features in the batch to form the similarity matrix $\mathbf{S} \in \mathbb{R}^{B \times B}$, where each element is defined as:
\begin{equation}
\mathbf{S}_{ij} = \frac{v_i \cdot t_j}{||v_i|| ||t_j||},
\label{eq:sim_cosine}
\end{equation}
in which $\mathbf{S}_{ij}$ denotes the similarity between $i$-th visual feature and the $j$-th textual feature. The soft-label supervision matrix $\mathbf{\tilde{A}} \in \mathbb{R}^{B \times B}$ is utilized to supervise the contrastive learning process. Each element $\mathbf{\tilde{A}}_{ij}$ reflects the semantic similarity between the input cases $i$ and $j$, enabling fine-grained alignment beyond one-hot label. Then, the contrastive loss is computed using a symmetrized cross-entropy formulation between visual and textual modalities:
\begin{equation}
\mathcal{L}_{\text{con}} = \frac{1}{2} \left( \text{CE}(\text{Softmax}(\mathbf{S}/\tau), \mathbf{\tilde{A}}) + \text{CE}(\text{Softmax}(\mathbf{S}^\top/\tau), \mathbf{\tilde{A}}^\top) \right)
\label{eq:contrastive_loss}
\end{equation}
where $\tau$ is the temperature hyper-parameter that scales the logits, and $\text{CE}(\cdot, \cdot)$ denotes the cross-entropy loss between the predicted similarities and the soft-label targets. In this way, Eq.~\ref{eq:contrastive_loss} encourages semantically similar image–text pairs to be pulled closer in the embedding space while pushing dissimilar pairs apart, ultimately enhancing the capacity to capture fine-grained cardiovascular semantics across modalities.

\subsection{Implementation Details}
All experiments are conducted using PyTorch on a single NVIDIA A6000 GPU, encompassing the pre-training, fine-tuning, and inference stages. To accommodate 3D volumetric medical image, we adopt the standard 2D ViT-B/32 model \cite{dosovitskiy2020image} as the visual encoder by implementing 3D patch embeddings. For textual encoder, we adopt PubMedBERT \cite{devlin2019bert, gu2021domain}, configured with 768-dimensional hidden states.

During the first-stage pre-training, we adopt the MAE framework with a 75\% random patch masking ratio. The model is trained for 50 epochs using a batch size of 64 and the AdamW optimizer, with a base learning rate of $1 \times 10^{-4}$ and weight decay of 0.01. In the second-stage contrastive pre-training, the batch size is set to 16 to accommodate the additional memory overhead of the textual encoder. We use the AdamW optimizer with a base learning rate of $1 \times 10^{-5}$ and a higher learning rate of $5 \times 10^{-5}$ for the projection layers, with a weight decay of 0.01. This phase is trained for 10 epochs. Meanwhile, both stages adopt a consistent learning rate scheduling strategy consisting of linear warm-up followed by cosine decay, which facilitates effective parameter optimization throughout the pre-training process.

For the downstream fine-tuning process, a classification head is appended to the pre-trained visual encoder. The learning rate is set to $1 \times 10^{-5}$ for the encoder and $5 \times 10^{-5}$ for the classification layers. Notably, a similar learning rate adjustment strategy is employed as in the pre-training phase.

\section{Conclusion}
This study introduces Cardiac-CLIP, a multi-modal foundation model designed specifically for 3D cardiac CT images. The model is trained through a two-stage pre-training process, self-supervised representation learning using 3D MAE followed by contrastive alignment with radiology reports, which enables Cardiac-CLIP to effectively bridge visual and textual modalities in complex cardiovascular diagnostic tasks. We conduct extensive evaluations across multi-center datasets from 12 different institutions, which covers various tasks such as cardiovascular abnormality classification, cross-modal retrieval, and high-level clinical analysis. Cardiac-CLIP consistently outperforms existing models and demonstrates strong generalization across both internal and external datasets in zero-shot and fine-tuning settings. Particularly, our model shows great capacity in clinically important tasks, including prospective risk prediction for acute coronary syndrome, diagnosis of functional coronary stenosis and Coronary Artery Calcium Grading. These results highlight the potential to support complex clinical decision-making. Overall, our findings suggest that Cardiac-CLIP holds significant promise for advancing cardiovascular diagnostics and facilitating real-world clinical applications.

\section{Data and Code Availability}
In our experiments, public available dataset could be download from their corresponding website. Our codes and pretrained parameters can be found in \url{https://github.com/Sliver-g/Cardiac-CLIP}.

\clearpage

\bibliographystyle{unsrtnat}  
\bibliography{reference}

\begin{thebibliography}{60}
\providecommand{\natexlab}[1]{#1}
\providecommand{\url}[1]{\texttt{#1}}
\expandafter\ifx\csname urlstyle\endcsname\relax
  \providecommand{\doi}[1]{doi: #1}\else
  \providecommand{\doi}{doi: \begingroup \urlstyle{rm}\Url}\fi

\bibitem[Martin et~al.(2025)Martin, Aday, Allen, Almarzooq, Anderson, Arora, Avery, Baker-Smith, Bansal, Beaton, et~al.]{1_1_2025_Heart_disease_statistics}
Seth~S Martin, Aaron~W Aday, Norrina~B Allen, Zaid~I Almarzooq, Cheryl~AM Anderson, Pankaj Arora, Christy~L Avery, Carissa~M Baker-Smith, Nisha Bansal, Andrea~Z Beaton, et~al.
\newblock 2025 heart disease and stroke statistics: A report of us and global data from the american heart association.
\newblock \emph{Circulation}, 2025.

\bibitem[Wang et~al.(2024)Wang, Wang, Lu, Li, Wu, Yang, Cui, Xu, Song, Yang, He, Zhang, Zhang, Li, and Hu]{1_2_China2020_2030}
Runsi Wang, Yunfeng Wang, Jiapeng Lu, Yichong Li, Chaoqun Wu, Yang Yang, Jianlan Cui, Wei Xu, Lijuan Song, Hao Yang, Wenyan He, Yan Zhang, Xingyi Zhang, Xi~Li, and Shengshou Hu.
\newblock Forecasting cardiovascular disease risk and burden in china from 2020 to 2030: a simulation study based on a nationwide cohort.
\newblock \emph{Heart}, 111\penalty0 (5):\penalty0 205--211, December 2024.
\newblock ISSN 1468-201X.

\bibitem[Zeleznik et~al.(2021{\natexlab{a}})Zeleznik, Foldyna, Eslami, Weiss, Alexander, Taron, Parmar, Alvi, Banerji, Uno, et~al.]{zeleznik2021deep}
Roman Zeleznik, Borek Foldyna, Parastou Eslami, Jakob Weiss, Ivanov Alexander, Jana Taron, Chintan Parmar, Raza~M Alvi, Dahlia Banerji, Mio Uno, et~al.
\newblock Deep convolutional neural networks to predict cardiovascular risk from computed tomography.
\newblock \emph{Nature communications}, 12\penalty0 (1):\penalty0 715, 2021{\natexlab{a}}.

\bibitem[Fairbairn et~al.(2025)Fairbairn, Mullen, Nicol, Lip, Schmitt, Shaw, Tidbury, Kemp, Crooks, Burnside, et~al.]{fairbairn2025implementation}
Timothy~A Fairbairn, Liam Mullen, Edward Nicol, Gregory~YH Lip, Matthias Schmitt, Matthew Shaw, Laurence Tidbury, Ian Kemp, Jennifer Crooks, Girvan Burnside, et~al.
\newblock Implementation of a national ai technology program on cardiovascular outcomes and the health system.
\newblock \emph{Nature Medicine}, pages 1--8, 2025.

\bibitem[Khera et~al.(2024)Khera, Oikonomou, Nadkarni, Morley, Wiens, Butte, and Topol]{3_1_AI4CVD_care_review}
Rohan Khera, Evangelos~K. Oikonomou, Girish~N. Nadkarni, Jessica~R. Morley, Jenna Wiens, Atul~J. Butte, and Eric~J. Topol.
\newblock Transforming cardiovascular care with artificial intelligence: From discovery to practice.
\newblock \emph{Journal of the American College of Cardiology}, 84\penalty0 (1):\penalty0 97--114, July 2024.
\newblock ISSN 0735-1097.

\bibitem[Wolterink et~al.(2015)Wolterink, Leiner, Takx, Viergever, and Isgum]{3_2}
Jelmer~M. Wolterink, Tim Leiner, Richard A.~P. Takx, Max~A. Viergever, and Ivana Isgum.
\newblock Automatic coronary calcium scoring in non-contrast-enhanced ecg-triggered cardiac ct with ambiguity detection.
\newblock \emph{IEEE Transactions on Medical Imaging}, 34\penalty0 (9):\penalty0 1867--1878, September 2015.
\newblock ISSN 1558-254X.

\bibitem[Yang et~al.(2016)Yang, Chen, Ning, Sun, Shu, and Coatrieux]{3_3Yang2016}
Guanyu Yang, Yang Chen, Xiufang Ning, Qiaoyu Sun, Huazhong Shu, and Jean-Louis Coatrieux.
\newblock Automatic coronary calcium scoring using noncontrast and contrast ct images: Automatic coronary calcium scoring.
\newblock \emph{Medical Physics}, 43\penalty0 (5):\penalty0 2174--2186, April 2016.
\newblock ISSN 0094-2405.

\bibitem[Wolterink et~al.(2016)Wolterink, Leiner, de~Vos, van Hamersvelt, Viergever, and Išgum]{3_4Wolterink2016}
Jelmer~M. Wolterink, Tim Leiner, Bob~D. de~Vos, Robbert~W. van Hamersvelt, Max~A. Viergever, and Ivana Išgum.
\newblock Automatic coronary artery calcium scoring in cardiac ct angiography using paired convolutional neural networks.
\newblock \emph{Medical Image Analysis}, 34:\penalty0 123--136, December 2016.
\newblock ISSN 1361-8415.

\bibitem[Zeleznik et~al.(2021{\natexlab{b}})Zeleznik, Foldyna, Eslami, Weiss, Alexander, Taron, Parmar, Alvi, Banerji, Uno, Kikuchi, Karady, Zhang, Scholtz, Mayrhofer, Lyass, Mahoney, Massaro, Vasan, Douglas, Hoffmann, Lu, and Aerts]{3_5_NC_Zeleznik2021}
Roman Zeleznik, Borek Foldyna, Parastou Eslami, Jakob Weiss, Ivanov Alexander, Jana Taron, Chintan Parmar, Raza~M. Alvi, Dahlia Banerji, Mio Uno, Yasuka Kikuchi, Julia Karady, Lili Zhang, Jan-Erik Scholtz, Thomas Mayrhofer, Asya Lyass, Taylor~F. Mahoney, Joseph~M. Massaro, Ramachandran~S. Vasan, Pamela~S. Douglas, Udo Hoffmann, Michael~T. Lu, and Hugo J. W.~L. Aerts.
\newblock Deep convolutional neural networks to predict cardiovascular risk from computed tomography.
\newblock \emph{Nature Communications}, 12\penalty0 (1), January 2021{\natexlab{b}}.
\newblock ISSN 2041-1723.

\bibitem[Eng et~al.(2021)Eng, Chute, Khandwala, Rajpurkar, Long, Shleifer, Khalaf, Sandhu, Rodriguez, Maron, Seyyedi, Marin, Golub, Budoff, Kitamura, Takahashi, Filice, Shah, Mongan, Kallianos, Langlotz, Lungren, Ng, and Patel]{3_6Eng2021}
David Eng, Christopher Chute, Nishith Khandwala, Pranav Rajpurkar, Jin Long, Sam Shleifer, Mohamed~H. Khalaf, Alexander~T. Sandhu, Fatima Rodriguez, David~J. Maron, Saeed Seyyedi, Daniele Marin, Ilana Golub, Matthew Budoff, Felipe Kitamura, Marcelo~Straus Takahashi, Ross~W. Filice, Rajesh Shah, John Mongan, Kimberly Kallianos, Curtis~P. Langlotz, Matthew~P. Lungren, Andrew~Y. Ng, and Bhavik~N. Patel.
\newblock Automated coronary calcium scoring using deep learning with multicenter external validation.
\newblock \emph{npj Digital Medicine}, 4\penalty0 (1), June 2021.
\newblock ISSN 2398-6352.

\bibitem[Zreik et~al.(2019{\natexlab{a}})Zreik, van Hamersvelt, Wolterink, Leiner, Viergever, and Isgum]{3_8Zreik2019}
Majd Zreik, Robbert~W. van Hamersvelt, Jelmer~M. Wolterink, Tim Leiner, Max~A. Viergever, and Ivana Isgum.
\newblock A recurrent cnn for automatic detection and classification of coronary artery plaque and stenosis in coronary ct angiography.
\newblock \emph{IEEE Transactions on Medical Imaging}, 38\penalty0 (7):\penalty0 1588--1598, July 2019{\natexlab{a}}.
\newblock ISSN 1558-254X.

\bibitem[Chao et~al.(2021)Chao, Shan, Homayounieh, Singh, Khera, Guo, Su, Wang, Kalra, and Yan]{3_9_NC_Chao2021}
Hanqing Chao, Hongming Shan, Fatemeh Homayounieh, Ramandeep Singh, Ruhani~Doda Khera, Hengtao Guo, Timothy Su, Ge~Wang, Mannudeep~K. Kalra, and Pingkun Yan.
\newblock Deep learning predicts cardiovascular disease risks from lung cancer screening low dose computed tomography.
\newblock \emph{Nature Communications}, 12\penalty0 (1), May 2021.
\newblock ISSN 2041-1723.

\bibitem[Zreik et~al.(2019{\natexlab{b}})Zreik, Van~Hamersvelt, Khalili, Wolterink, Voskuil, Viergever, Leiner, and I{\v{s}}gum]{zreik2019deep}
Majd Zreik, Robbert~W Van~Hamersvelt, Nadieh Khalili, Jelmer~M Wolterink, Michiel Voskuil, Max~A Viergever, Tim Leiner, and Ivana I{\v{s}}gum.
\newblock Deep learning analysis of coronary arteries in cardiac ct angiography for detection of patients requiring invasive coronary angiography.
\newblock \emph{IEEE Transactions on Medical Imaging}, 39\penalty0 (5):\penalty0 1545--1557, 2019{\natexlab{b}}.

\bibitem[van Velzen et~al.(2020)van Velzen, Lessmann, Velthuis, Bank, van~den Bongard, Leiner, de~Jong, Veldhuis, Correa, Terry, et~al.]{van2020deep}
Sanne~GM van Velzen, Nikolas Lessmann, Birgitta~K Velthuis, Ingrid~EM Bank, Desiree~HJG van~den Bongard, Tim Leiner, Pim~A de~Jong, Wouter~B Veldhuis, Adolfo Correa, James~G Terry, et~al.
\newblock Deep learning for automatic calcium scoring in ct: validation using multiple cardiac ct and chest ct protocols.
\newblock \emph{Radiology}, 295\penalty0 (1):\penalty0 66--79, 2020.

\bibitem[Qi et~al.(2023)Qi, He, Qi, Zhang, and Yang]{qi2023dynamic}
Yaolei Qi, Yuting He, Xiaoming Qi, Yuan Zhang, and Guanyu Yang.
\newblock Dynamic snake convolution based on topological geometric constraints for tubular structure segmentation.
\newblock In \emph{Proceedings of the IEEE/CVF International Conference on Computer Vision}, pages 6070--6079, 2023.

\bibitem[Denzinger et~al.(2019)Denzinger, Wels, Ravikumar, Breininger, Reidelsh{\"o}fer, Eckert, S{\"u}hling, Schmermund, and Maier]{denzinger2019coronary}
Felix Denzinger, Michael Wels, Nishant Ravikumar, Katharina Breininger, Anika Reidelsh{\"o}fer, Joachim Eckert, Michael S{\"u}hling, Axel Schmermund, and Andreas Maier.
\newblock Coronary artery plaque characterization from ccta scans using deep learning and radiomics.
\newblock In \emph{Medical Image Computing and Computer Assisted Intervention--MICCAI 2019: 22nd International Conference, Shenzhen, China, October 13--17, 2019, Proceedings, Part IV 22}, pages 593--601. Springer, 2019.

\bibitem[Dey et~al.(2018)Dey, Gaur, Ovrehus, Slomka, Betancur, Goeller, Hell, Gransar, Berman, Achenbach, et~al.]{dey2018integrated}
Damini Dey, Sara Gaur, Kristian~A Ovrehus, Piotr~J Slomka, Julian Betancur, Markus Goeller, Michaela~M Hell, Heidi Gransar, Daniel~S Berman, Stephan Achenbach, et~al.
\newblock Integrated prediction of lesion-specific ischaemia from quantitative coronary ct angiography using machine learning: a multicentre study.
\newblock \emph{European Radiology}, 28:\penalty0 2655--2664, 2018.

\bibitem[Tizhoosh and Pantanowitz(2018)]{tizhoosh2018artificial}
Hamid~Reza Tizhoosh and Liron Pantanowitz.
\newblock Artificial intelligence and digital pathology: challenges and opportunities.
\newblock \emph{Journal of Pathology Informatics}, 9\penalty0 (1):\penalty0 38, 2018.

\bibitem[Wang et~al.(2021)Wang, Li, Wang, Liu, Wang, Tan, Wu, Liu, Sun, Yang, et~al.]{wang2021annotation}
Shanshan Wang, Cheng Li, Rongpin Wang, Zaiyi Liu, Meiyun Wang, Hongna Tan, Yaping Wu, Xinfeng Liu, Hui Sun, Rui Yang, et~al.
\newblock Annotation-efficient deep learning for automatic medical image segmentation.
\newblock \emph{Nature Communications}, 12\penalty0 (1):\penalty0 5915, 2021.

\bibitem[T{\"o}lle et~al.(2025)T{\"o}lle, Garthe, Scherer, Seliger, Leha, Kr{\"u}ger, Simm, Martin, Eble, Kelm, et~al.]{tolle2025real}
Malte T{\"o}lle, Philipp Garthe, Clemens Scherer, Jan~Moritz Seliger, Andreas Leha, Nina Kr{\"u}ger, Stefan Simm, Simon Martin, Sebastian Eble, Halvar Kelm, et~al.
\newblock Real world federated learning with a knowledge distilled transformer for cardiac ct imaging.
\newblock \emph{npj Digital Medicine}, 8\penalty0 (1):\penalty0 88, 2025.

\bibitem[Guan and Liu(2021)]{guan2021domain}
Hao Guan and Mingxia Liu.
\newblock Domain adaptation for medical image analysis: a survey.
\newblock \emph{IEEE Transactions on Biomedical Engineering}, 69\penalty0 (3):\penalty0 1173--1185, 2021.

\bibitem[Radford et~al.(2021)Radford, Kim, Hallacy, Ramesh, Goh, Agarwal, Sastry, Askell, Mishkin, Clark, Krueger, and Sutskever]{4_2OrigCLIP}
Alec Radford, Jong~Wook Kim, Chris Hallacy, Aditya Ramesh, Gabriel Goh, Sandhini Agarwal, Girish Sastry, Amanda Askell, Pamela Mishkin, Jack Clark, Gretchen Krueger, and Ilya Sutskever.
\newblock Learning transferable visual models from natural language supervision.
\newblock In \emph{International Conference on Machine Learning}, 2021.

\bibitem[Jia et~al.(2021)Jia, Yang, Xia, Chen, Parekh, Pham, Le, Sung, Li, and Duerig]{jia2021scaling}
Chao Jia, Yinfei Yang, Ye~Xia, Yi-Ting Chen, Zarana Parekh, Hieu Pham, Quoc Le, Yun-Hsuan Sung, Zhen Li, and Tom Duerig.
\newblock Scaling up visual and vision-language representation learning with noisy text supervision.
\newblock In \emph{International Conference on Machine Learning}, pages 4904--4916. PMLR, 2021.

\bibitem[Singh et~al.(2022)Singh, Hu, Goswami, Couairon, Galuba, Rohrbach, and Kiela]{singh2022flava}
Amanpreet Singh, Ronghang Hu, Vedanuj Goswami, Guillaume Couairon, Wojciech Galuba, Marcus Rohrbach, and Douwe Kiela.
\newblock Flava: A foundational language and vision alignment model.
\newblock In \emph{Proceedings of the IEEE/CVF Conference on Computer Vision and Pattern Recognition}, pages 15638--15650, 2022.

\bibitem[Xiang et~al.(2025)Xiang, Wang, Zhang, Xi, Eweje, Chen, Li, Bergstrom, Gopaulchan, Kim, Yu, Willens, Olguin, Nirschl, Neal, Diehn, Yang, and Li]{4_3_N_pathology_Xiang2025}
Jinxi Xiang, Xiyue Wang, Xiaoming Zhang, Yinghua Xi, Feyisope Eweje, Yijiang Chen, Yuchen Li, Colin Bergstrom, Matthew Gopaulchan, Ted Kim, Kun-Hsing Yu, Sierra Willens, Francesca~Maria Olguin, Jeffrey~J. Nirschl, Joel Neal, Maximilian Diehn, Sen Yang, and Ruijiang Li.
\newblock A vision–language foundation model for precision oncology.
\newblock \emph{Nature}, January 2025.
\newblock ISSN 1476-4687.

\bibitem[Lu et~al.(2024)Lu, Chen, Williamson, Chen, Liang, Ding, Jaume, Odintsov, Le, Gerber, Parwani, Zhang, and Mahmood]{4_6_NM_pathology_Lu2024}
Ming~Y. Lu, Bowen Chen, Drew F.~K. Williamson, Richard~J. Chen, Ivy Liang, Tong Ding, Guillaume Jaume, Igor Odintsov, Long~Phi Le, Georg Gerber, Anil~V. Parwani, Andrew Zhang, and Faisal Mahmood.
\newblock A visual-language foundation model for computational pathology.
\newblock \emph{Nature Medicine}, 30\penalty0 (3):\penalty0 863--874, March 2024.
\newblock ISSN 1546-170X.

\bibitem[Huang et~al.(2023)Huang, Bianchi, Yuksekgonul, Montine, and Zou]{4_7_NM_pathology_Huang2023}
Zhi Huang, Federico Bianchi, Mert Yuksekgonul, Thomas~J. Montine, and James Zou.
\newblock A visual–language foundation model for pathology image analysis using medical twitter.
\newblock \emph{Nature Medicine}, 29\penalty0 (9):\penalty0 2307--2316, August 2023.
\newblock ISSN 1546-170X.

\bibitem[Zhang et~al.(2023)Zhang, Wu, Zhang, Xie, and Wang]{4_8_KnowledgeEn_NC_X_Zhang2023}
Xiaoman Zhang, Chaoyi Wu, Ya~Zhang, Weidi Xie, and Yanfeng Wang.
\newblock Knowledge-enhanced visual-language pre-training on chest radiology images.
\newblock \emph{Nature Communications}, 14\penalty0 (1), July 2023.
\newblock ISSN 2041-1723.

\bibitem[Huang et~al.(2024)Huang, Li, Zhou, Yang, Liu, Liang, Zheng, Zhang, and Wang]{4_9_MAE_NC_X_Huang2024}
Weijian Huang, Cheng Li, Hong-Yu Zhou, Hao Yang, Jiarun Liu, Yong Liang, Hairong Zheng, Shaoting Zhang, and Shanshan Wang.
\newblock Enhancing representation in radiography-reports foundation model: a granular alignment algorithm using masked contrastive learning.
\newblock \emph{Nature Communications}, 15\penalty0 (1), September 2024.
\newblock ISSN 2041-1723.

\bibitem[Zhou et~al.(2022)Zhou, Chen, Zhang, Luo, Wang, and Yu]{4_11_NMI_X_Zhou2022}
Hong-Yu Zhou, Xiaoyu Chen, Yinghao Zhang, Ruibang Luo, Liansheng Wang, and Yizhou Yu.
\newblock Generalized radiograph representation learning via cross-supervision between images and free-text radiology reports.
\newblock \emph{Nature Machine Intelligence}, 4\penalty0 (1):\penalty0 32--40, January 2022.
\newblock ISSN 2522-5839.

\bibitem[Wu et~al.(2023)Wu, Zhang, Zhang, Wang, and Xie]{4_15_KnowledgeEn_X_Wu_2023_ICCV}
Chaoyi Wu, Xiaoman Zhang, Ya~Zhang, Yanfeng Wang, and Weidi Xie.
\newblock Medklip: Medical knowledge enhanced language-image pre-training for x-ray diagnosis.
\newblock In \emph{Proceedings of the IEEE/CVF International Conference on Computer Vision}, pages 21372--21383, October 2023.

\bibitem[Lai et~al.(2024)Lai, Yao, Jiang, Wang, He, Tao, and Zhou]{4_16_prompt_soft_X_Lai2024}
Haoran Lai, Qingsong Yao, Zihang Jiang, Rongsheng Wang, Zhiyang He, Xiaodong Tao, and S.~Kevin Zhou.
\newblock Carzero: Cross-attention alignment for radiology zero-shot classification.
\newblock In \emph{Proceedings of the IEEE/CVF Conference on Computer Vision and Pattern Recognition}, pages 11137--11146. IEEE, June 2024.

\bibitem[Wang et~al.(2022)Wang, Wu, Agarwal, and Sun]{4_17_soft_X_Wang2022}
Zifeng Wang, Zhenbang Wu, Dinesh Agarwal, and Jimeng Sun.
\newblock Medclip: Contrastive learning from unpaired medical images and text.
\newblock In \emph{Proceedings of the 2022 Conference on Empirical Methods in Natural Language Processing}. Association for Computational Linguistics, 2022.

\bibitem[Kim et~al.(2024)Kim, Gadgil, DeGrave, Omiye, Cai, Daneshjou, and Lee]{4_12_NM_skin_Kim2024}
Chanwoo Kim, Soham~U. Gadgil, Alex~J. DeGrave, Jesutofunmi~A. Omiye, Zhuo~Ran Cai, Roxana Daneshjou, and Su-In Lee.
\newblock Transparent medical image ai via an image–text foundation model grounded in medical literature.
\newblock \emph{Nature Medicine}, 30\penalty0 (4):\penalty0 1154--1165, April 2024.
\newblock ISSN 1546-170X.

\bibitem[Zhou et~al.(2024)Zhou, He, Sun, Xu, Chen, Chu, Zhou, Liao, Zhang, Afvari, and Gao]{4_13_NC_skin_Zhou2024}
Juexiao Zhou, Xiaonan He, Liyuan Sun, Jiannan Xu, Xiuying Chen, Yuetan Chu, Longxi Zhou, Xingyu Liao, Bin Zhang, Shawn Afvari, and Xin Gao.
\newblock Pre-trained multimodal large language model enhances dermatological diagnosis using skingpt-4.
\newblock \emph{Nature Communications}, 15\penalty0 (1), July 2024.
\newblock ISSN 2041-1723.

\bibitem[Blankemeier et~al.(2024)Blankemeier, Cohen, Kumar, Van~Veen, Gardezi, Paschali, Chen, Delbrouck, Reis, Truyts, et~al.]{5_4_blankemeier2024merlin}
Louis Blankemeier, Joseph~Paul Cohen, Ashwin Kumar, Dave Van~Veen, Syed Jamal~Safdar Gardezi, Magdalini Paschali, Zhihong Chen, Jean-Benoit Delbrouck, Eduardo Reis, Cesar Truyts, et~al.
\newblock Merlin: A vision language foundation model for 3d computed tomography.
\newblock \emph{Research Square}, pages rs--3, 2024.

\bibitem[Hamamci et~al.(2024)Hamamci, Er, Almas, Simsek, Esirgun, Dogan, Dasdelen, Durugol, Wittmann, Amiranashvili, Simsar, Simsar, Erdemir, Alanbay, Sekuboyina, Lafci, Bluethgen, Ozdemir, and Menze]{5_5_Hamamci2024}
Ibrahim~Ethem Hamamci, Sezgin Er, Furkan Almas, Ayse~Gulnihan Simsek, Sevval~Nil Esirgun, Irem Dogan, Muhammed~Furkan Dasdelen, Omer~Faruk Durugol, Bastian Wittmann, Tamaz Amiranashvili, Enis Simsar, Mehmet Simsar, Emine~Bensu Erdemir, Abdullah Alanbay, Anjany Sekuboyina, Berkan Lafci, Christian Bluethgen, Mehmet~Kemal Ozdemir, and Bjoern Menze.
\newblock Developing generalist foundation models from a multimodal dataset for 3d computed tomography, 2024.

\bibitem[Bai et~al.(2024)Bai, Du, Huang, Meng, and Zhao]{5_6_bai2024m3d}
Fan Bai, Yuxin Du, Tiejun Huang, Max Q-H Meng, and Bo~Zhao.
\newblock M3d: Advancing 3d medical image analysis with multi-modal large language models.
\newblock \emph{arXiv preprint arXiv:2404.00578}, 2024.

\bibitem[Cao et~al.(2024)Cao, Zhang, Xia, Mok, Li, Ye, Lu, Zheng, Tang, and Zhang]{5_1_soft_CHestCT_Cao2024}
Weiwei Cao, Jianpeng Zhang, Yingda Xia, Tony C.~W. Mok, Zi~Li, Xianghua Ye, Le~Lu, Jian Zheng, Yuxing Tang, and Ling Zhang.
\newblock Bootstrapping chest ct image understanding by distilling knowledge from x-ray expert models.
\newblock In \emph{Proceedings of the IEEE/CVF Conference on Computer Vision and Pattern Recognition}, pages 11238--11247. IEEE, June 2024.

\bibitem[Lin et~al.(2023)Lin, Zhao, Zhang, Wu, Zhang, Wang, and Xie]{lin2023pmc}
Weixiong Lin, Ziheng Zhao, Xiaoman Zhang, Chaoyi Wu, Ya~Zhang, Yanfeng Wang, and Weidi Xie.
\newblock Pmc-clip: Contrastive language-image pre-training using biomedical documents.
\newblock In \emph{International Conference on Medical Image Computing and Computer-Assisted Intervention}, pages 525--536. Springer, 2023.

\bibitem[Chen et~al.(2022)Chen, Du, Hu, Liu, Li, Wan, and Chang]{8_2_M3AE_Chen2022}
Zhihong Chen, Yuhao Du, Jinpeng Hu, Yang Liu, Guanbin Li, Xiang Wan, and Tsung-Hui Chang.
\newblock \emph{Multi-modal Masked Autoencoders for Medical Vision-and-Language Pre-training}, pages 679--689.
\newblock Springer Nature Switzerland, 2022.
\newblock ISBN 9783031164439.

\bibitem[Zhou et~al.(2023)Zhou, Chia, Wagner, Ayhan, Williamson, Struyven, Liu, Xu, Lozano, Woodward-Court, Kihara, Allen, Gallacher, Littlejohns, Aslam, Bishop, Black, Sergouniotis, Atan, Dick, Williams, Barman, Barrett, Mackie, Braithwaite, Carare, Ennis, Gibson, Lotery, Self, Chakravarthy, Hogg, Paterson, Woodside, Peto, Mckay, Mcguinness, Foster, Balaskas, Khawaja, Pontikos, Rahi, Lascaratos, Patel, Chan, Chua, Day, Desai, Egan, Fruttiger, Garway-Heath, Hardcastle, Khaw, Moore, Sivaprasad, Strouthidis, Thomas, Tufail, Viswanathan, Dhillon, Macgillivray, Sudlow, Vitart, Doney, Trucco, Guggeinheim, Morgan, Hammond, Williams, Hysi, Harding, Zheng, Luben, Luthert, Sun, McKibbin, O’Sullivan, Oram, Weedon, Owen, Rudnicka, Sattar, Steel, Stratton, Tapp, Yates, Petzold, Madhusudhan, Altmann, Lee, Topol, Denniston, Alexander, and Keane]{4_14_N_retinal_Zhou2023}
Yukun Zhou, Mark~A. Chia, Siegfried~K. Wagner, Murat~S. Ayhan, Dominic~J. Williamson, Robbert~R. Struyven, Timing Liu, Moucheng Xu, Mateo~G. Lozano, Peter Woodward-Court, Yuka Kihara, Naomi Allen, John E.~J. Gallacher, Thomas Littlejohns, Tariq Aslam, Paul Bishop, Graeme Black, Panagiotis Sergouniotis, Denize Atan, Andrew~D. Dick, Cathy Williams, Sarah Barman, Jenny~H. Barrett, Sarah Mackie, Tasanee Braithwaite, Roxana~O. Carare, Sarah Ennis, Jane Gibson, Andrew~J. Lotery, Jay Self, Usha Chakravarthy, Ruth~E. Hogg, Euan Paterson, Jayne Woodside, Tunde Peto, Gareth Mckay, Bernadette Mcguinness, Paul~J. Foster, Konstantinos Balaskas, Anthony~P. Khawaja, Nikolas Pontikos, Jugnoo~S. Rahi, Gerassimos Lascaratos, Praveen~J. Patel, Michelle Chan, Sharon Y.~L. Chua, Alexander Day, Parul Desai, Cathy Egan, Marcus Fruttiger, David~F. Garway-Heath, Alison Hardcastle, Sir Peng~T. Khaw, Tony Moore, Sobha Sivaprasad, Nicholas Strouthidis, Dhanes Thomas, Adnan Tufail, Ananth~C. Viswanathan, Bal Dhillon, Tom Macgillivray,
  Cathie Sudlow, Veronique Vitart, Alexander Doney, Emanuele Trucco, Jeremy~A. Guggeinheim, James~E. Morgan, Chris~J. Hammond, Katie Williams, Pirro Hysi, Simon~P. Harding, Yalin Zheng, Robert Luben, Phil Luthert, Zihan Sun, Martin McKibbin, Eoin O’Sullivan, Richard Oram, Mike Weedon, Chris~G. Owen, Alicja~R. Rudnicka, Naveed Sattar, David Steel, Irene Stratton, Robyn Tapp, Max~M. Yates, Axel Petzold, Savita Madhusudhan, Andre Altmann, Aaron~Y. Lee, Eric~J. Topol, Alastair~K. Denniston, Daniel~C. Alexander, and Pearse~A. Keane.
\newblock A foundation model for generalizable disease detection from retinal images.
\newblock \emph{Nature}, 622\penalty0 (7981):\penalty0 156--163, September 2023.
\newblock ISSN 1476-4687.

\bibitem[null null(2011)]{7_NLSTdata_2011}
null null.
\newblock Reduced lung-cancer mortality with low-dose computed tomographic screening.
\newblock \emph{New England Journal of Medicine}, 365\penalty0 (5):\penalty0 395--409, August 2011.
\newblock ISSN 1533-4406.

\bibitem[He et~al.(2022)He, Chen, Xie, Li, Dollar, and Girshick]{8_MAE_He2022}
Kaiming He, Xinlei Chen, Saining Xie, Yanghao Li, Piotr Dollar, and Ross Girshick.
\newblock Masked autoencoders are scalable vision learners.
\newblock In \emph{Proceedings of the IEEE/CVF Conference on Computer Vision and Pattern Recognition}. IEEE, June 2022.

\bibitem[Zhang et~al.(2021)Zhang, Bao, and Ma]{zhang2021self}
Linfeng Zhang, Chenglong Bao, and Kaisheng Ma.
\newblock Self-distillation: Towards efficient and compact neural networks.
\newblock \emph{IEEE Transactions on Pattern Analysis and Machine Intelligence}, 44\penalty0 (8):\penalty0 4388--4403, 2021.

\bibitem[Wang and Yoon(2021)]{wang2021knowledge}
Lin Wang and Kuk-Jin Yoon.
\newblock Knowledge distillation and student-teacher learning for visual intelligence: A review and new outlooks.
\newblock \emph{IEEE Transactions on Pattern Analysis and Machine Intelligence}, 44\penalty0 (6):\penalty0 3048--3068, 2021.

\bibitem[Hu et~al.(2021)Hu, Jiang, Liu, Luo, Hu, Cao, Zhang, and Zhang]{hu2021hierarchical}
Yutao Hu, Xiaolong Jiang, Xuhui Liu, Xiaoyan Luo, Yao Hu, Xianbin Cao, Baochang Zhang, and Jun Zhang.
\newblock Hierarchical self-distilled feature learning for fine-grained visual categorization.
\newblock \emph{IEEE Transactions on Neural Networks and Learning Systems}, 2021.

\bibitem[Bergmark et~al.(2022)Bergmark, Mathenge, Merlini, Lawrence-Wright, and Giugliano]{bergmark2022acute}
Brian~A Bergmark, Njambi Mathenge, Piera~A Merlini, Marilyn~B Lawrence-Wright, and Robert~P Giugliano.
\newblock Acute coronary syndromes.
\newblock \emph{The Lancet}, 399\penalty0 (10332):\penalty0 1347--1358, 2022.

\bibitem[Al-Zaiti et~al.(2020)Al-Zaiti, Besomi, Bouzid, Faramand, Frisch, Martin-Gill, Gregg, Saba, Callaway, and Sejdi{\'c}]{al2020machine}
Salah Al-Zaiti, Lucas Besomi, Zeineb Bouzid, Ziad Faramand, Stephanie Frisch, Christian Martin-Gill, Richard Gregg, Samir Saba, Clifton Callaway, and Ervin Sejdi{\'c}.
\newblock Machine learning-based prediction of acute coronary syndrome using only the pre-hospital 12-lead electrocardiogram.
\newblock \emph{Nature Communications}, 11\penalty0 (1):\penalty0 3966, 2020.

\bibitem[Ko et~al.(2017)Ko, Cameron, Munnur, Wong, Fujisawa, Sakaguchi, Hirohata, Hislop-Jambrich, Fujimoto, Takamura, et~al.]{ko2017noninvasive}
Brian~S Ko, James~D Cameron, Ravi~K Munnur, Dennis~TL Wong, Yasuko Fujisawa, Takuya Sakaguchi, Kenji Hirohata, Jacqui Hislop-Jambrich, Shinichiro Fujimoto, Kazuhisa Takamura, et~al.
\newblock Noninvasive ct-derived ffr based on structural and fluid analysis: a comparison with invasive ffr for detection of functionally significant stenosis.
\newblock \emph{JACC: Cardiovascular Imaging}, 10\penalty0 (6):\penalty0 663--673, 2017.

\bibitem[Hinton et~al.(2015)Hinton, Vinyals, and Dean]{hinton2015distilling}
Geoffrey Hinton, Oriol Vinyals, and Jeff Dean.
\newblock Distilling the knowledge in a neural network.
\newblock \emph{arXiv preprint arXiv:1503.02531}, 2015.

\bibitem[Xu et~al.(2020)Xu, Liu, Li, and Loy]{xu2020knowledge}
Guodong Xu, Ziwei Liu, Xiaoxiao Li, and Chen~Change Loy.
\newblock Knowledge distillation meets self-supervision.
\newblock In \emph{Proceedings of the European Conference on Computer Vision}, pages 588--604. Springer, 2020.

\bibitem[Min et~al.(2012)Min, Leipsic, Pencina, Berman, Koo, Van~Mieghem, Erglis, Lin, Dunning, Apruzzese, et~al.]{min2012diagnostic}
James~K Min, Jonathon Leipsic, Michael~J Pencina, Daniel~S Berman, Bon-Kwon Koo, Carlos Van~Mieghem, Andrejs Erglis, Fay~Y Lin, Allison~M Dunning, Patricia Apruzzese, et~al.
\newblock Diagnostic accuracy of fractional flow reserve from anatomic ct angiography.
\newblock \emph{Jama}, 308\penalty0 (12):\penalty0 1237--1245, 2012.

\bibitem[Agatston et~al.(1990)Agatston, Janowitz, Hildner, Zusmer, Viamonte~Jr, and Detrano]{agatston1990quantification}
Arthur~S Agatston, Warren~R Janowitz, Frank~J Hildner, Noel~R Zusmer, Manuel Viamonte~Jr, and Robert Detrano.
\newblock Quantification of coronary artery calcium using ultrafast computed tomography.
\newblock \emph{Journal of the American College of Cardiology}, 15\penalty0 (4):\penalty0 827--832, 1990.

\bibitem[Hecht et~al.(2017)Hecht, Cronin, Blaha, Budoff, Kazerooni, Narula, Yankelevitz, and Abbara]{hecht20172016}
Harvey~S Hecht, Paul Cronin, Michael~J Blaha, Matthew~J Budoff, Ella~A Kazerooni, Jagat Narula, David Yankelevitz, and Suhny Abbara.
\newblock 2016 scct/str guidelines for coronary artery calcium scoring of noncontrast noncardiac chest ct scans: a report of the society of cardiovascular computed tomography and society of thoracic radiology.
\newblock \emph{Journal of cardiovascular Computed Tomography}, 11\penalty0 (1):\penalty0 74--84, 2017.

\bibitem[Chefer et~al.(2021)Chefer, Gur, and Wolf]{chefer2021generic}
Hila Chefer, Shir Gur, and Lior Wolf.
\newblock Generic attention-model explainability for interpreting bi-modal and encoder-decoder transformers.
\newblock In \emph{Proceedings of the IEEE/CVF international conference on computer vision}, pages 397--406, 2021.

\bibitem[Hurst et~al.(2024)Hurst, Lerer, Goucher, Perelman, Ramesh, Clark, Ostrow, Welihinda, Hayes, Radford, et~al.]{hurst2024gpt}
Aaron Hurst, Adam Lerer, Adam~P Goucher, Adam Perelman, Aditya Ramesh, Aidan Clark, AJ~Ostrow, Akila Welihinda, Alan Hayes, Alec Radford, et~al.
\newblock Gpt-4o system card.
\newblock \emph{arXiv preprint arXiv:2410.21276}, 2024.

\bibitem[Dosovitskiy et~al.(2020)Dosovitskiy, Beyer, Kolesnikov, Weissenborn, Zhai, Unterthiner, Dehghani, Minderer, Heigold, Gelly, et~al.]{dosovitskiy2020image}
Alexey Dosovitskiy, Lucas Beyer, Alexander Kolesnikov, Dirk Weissenborn, Xiaohua Zhai, Thomas Unterthiner, Mostafa Dehghani, Matthias Minderer, Georg Heigold, Sylvain Gelly, et~al.
\newblock An image is worth 16x16 words: Transformers for image recognition at scale.
\newblock \emph{arXiv preprint arXiv:2010.11929}, 2020.

\bibitem[Devlin et~al.(2019)Devlin, Chang, Lee, and Toutanova]{devlin2019bert}
Jacob Devlin, Ming-Wei Chang, Kenton Lee, and Kristina Toutanova.
\newblock Bert: Pre-training of deep bidirectional transformers for language understanding.
\newblock In \emph{Proceedings of the 2019 conference of the North American chapter of the association for computational linguistics: human language technologies, volume 1 (long and short papers)}, pages 4171--4186, 2019.

\bibitem[Gu et~al.(2021)Gu, Tinn, Cheng, Lucas, Usuyama, Liu, Naumann, Gao, and Poon]{gu2021domain}
Yu~Gu, Robert Tinn, Hao Cheng, Michael Lucas, Naoto Usuyama, Xiaodong Liu, Tristan Naumann, Jianfeng Gao, and Hoifung Poon.
\newblock Domain-specific language model pretraining for biomedical natural language processing.
\newblock \emph{ACM Transactions on Computing for Healthcare}, 3\penalty0 (1):\penalty0 1--23, 2021.

\end{thebibliography}

\end{document}


\captionsetup[figure]{name=Supplementary Fig.}
\captionsetup[table]{name=Supplementary Table }

In this chapter, we assess diagnostic tasks at two levels of granularity: ICD-10-CM and individual disorders. To achieve this, we employ both late fusion and early fusion strategies. Specifically, in the final fusion stage of the early fusion approach, the measurement options include: 

\begin{itemize}
    \item \textbf{Random Pick.} We randomly select an image of a specific modality from a view within a case. It is grounded in the belief that each view of a case should encompass the essential information required for an accurate 
    \item \textbf{Max on Class.} We choose the maximum score in each classification class among all the images in a case. It is founded on the principle that if there is an image within a case that can potentially illustrate the presence of a specific disease, we should consider it essential to diagnose that disease.
    \item  \textbf{Mean on Class.} We calculate the mean score in each classification class among all the images in a case. It is rooted in the notion that relying solely on the diagnosis from a single image is insufficient, and our aim is to collectively evaluate all the images within a case for a comprehensive diagnosis.
\end{itemize}

Subsequently, we will present a comparative analysis of the outcomes obtained through these strategies, focusing on the two classification standards, ICD-10-CM and disorders.

\begin{table}[htbp]
\centering
\small
\setlength{\tabcolsep}{8pt}
\caption{Results on ICD10-CM~\chaoyi{change early/late fusion words.Add class number}}
\vspace{3pt}
\begin{tabular}{cccccccccc}
\hline
\multirow{2}{*}{\textbf{Granularity}}            & \multirow{2}{*}{\textbf{Split}}              & \multirow{2}{*}{\textbf{Fusion Mode}} & \multicolumn{7}{c}{\textbf{Metrics}}                                                                                 \\
                                                 &                                              &                                       & AUC            & AP             & F1             & MCC            & R@0.01         & R@0.05         & R@0.1          \\ \hline
\multicolumn{1}{c|}{\multirow{12}{*}{ICD-10 CM}} & \multicolumn{1}{c|}{\multirow{4}{*}{Head}}   & \multicolumn{1}{c|}{Late Fusion Rand} & 89.12          & 12.39          & 20.51          & 21.72          & 24.12          & 51.82          & 67.93          \\
\multicolumn{1}{c|}{}                            & \multicolumn{1}{c|}{}                        & \multicolumn{1}{c|}{Late Fusion Max}  & 90.39          & 12.80          & 20.65          & 21.94          & 24.39          & 54.12          & 70.11          \\
\multicolumn{1}{c|}{}                            & \multicolumn{1}{c|}{}                        & \multicolumn{1}{c|}{Late Fusion Mean} & 85.49          & 13.05          & 20.25          & 21.52          & 24.31          & 49.92          & 65.28          \\
\multicolumn{1}{c|}{}                            & \multicolumn{1}{c|}{}                        & \multicolumn{1}{c|}{Early Fusion}     & \textbf{91.15} & \textbf{14.38} & \textbf{22.83} & \textbf{24.12} & \textbf{28.39} & \textbf{58.82} & \textbf{73.48} \\ \cline{2-10} 
\multicolumn{1}{c|}{}                            & \multicolumn{1}{c|}{\multirow{4}{*}{Medium}} & \multicolumn{1}{c|}{Late Fusion Rand} & 88.98          & 7.72           & 15.44          & 18.02          & 22.52          & 50.50          & 63.96          \\
\multicolumn{1}{c|}{}                            & \multicolumn{1}{c|}{}                        & \multicolumn{1}{c|}{Late Fusion Max}  & \textbf{90.15} & 7.74           & 15.41          & 17.96          & 22.75          & 51.10          & 66.31          \\
\multicolumn{1}{c|}{}                            & \multicolumn{1}{c|}{}                        & \multicolumn{1}{c|}{Late Fusion Mean} & 85.05          & 7.69           & 15.03          & 17.53          & 22.58          & 47.41          & 64.52          \\
\multicolumn{1}{c|}{}                            & \multicolumn{1}{c|}{}                        & \multicolumn{1}{c|}{Early Fusion}     & 89.92          & \textbf{9.24}  & \textbf{17.14} & \textbf{19.86} & \textbf{26.89} & \textbf{54.31} & \textbf{66.52} \\ \cline{2-10} 
\multicolumn{1}{c|}{}                            & \multicolumn{1}{c|}{\multirow{4}{*}{Tail}}   & \multicolumn{1}{c|}{Late Fusion Rand} & 86.41          & 4.46           & \textbf{8.54}  & \textbf{12.48} & 8.37           & 23.74          & 37.12          \\
\multicolumn{1}{c|}{}                            & \multicolumn{1}{c|}{}                        & \multicolumn{1}{c|}{Late Fusion Max}  & \textbf{86.46} & 2.90           & 6.09           & 10.27          & 8.51           & \textbf{25.06} & \textbf{42.80} \\
\multicolumn{1}{c|}{}                            & \multicolumn{1}{c|}{}                        & \multicolumn{1}{c|}{Late Fusion Mean} & 81.50          & 2.68           & 5.42           & 9.04           & 6.15           & 20.13          & 30.64          \\
\multicolumn{1}{c|}{}                            & \multicolumn{1}{c|}{}                        & \multicolumn{1}{c|}{Early Fusion}     & 83.87          & \textbf{4.27}  & 7.88           & 11.74          & \textbf{8.83}  & 22.69          & 36.01          \\ \hline
\end{tabular}
\end{table}

\begin{table}[htbp]
\centering
\small
\setlength{\tabcolsep}{8pt}
\caption{Results on Disorders}
\vspace{3pt}
\begin{tabular}{cccccccccc}
\hline
\multirow{2}{*}{\textbf{Granularity}}            & \multirow{2}{*}{\textbf{Split}} & \multirow{2}{*}{\textbf{Fusion Mode}} & \multicolumn{7}{c}{\textbf{Metrics}}                                                                                                    \\
                                                 &                                 &                                       & AUC            & AP                                & F1             & MCC            & R@0.01         & R@0.05         & R@0.1          \\ \hline
\multicolumn{1}{c|}{\multirow{12}{*}{Disorders}} & \multirow{4}{*}{Head}           & \multicolumn{1}{c|}{Late Fusion Rand} & 91.32          & 14.08                             & 22.93          & 24.61          & 30.68          & 58.78          & 73.79          \\
\multicolumn{1}{c|}{}                            &                                 & \multicolumn{1}{c|}{Late Fusion Max}  & 91.91          & 13.56                             & 22.08          & 23.66          & 30.88          & 60.50          & 74.70          \\
\multicolumn{1}{c|}{}                            &                                 & \multicolumn{1}{c|}{Late Fusion Mean} & 88.72          & \textbf{14.29}                    & 23.26          & 24.60          & 31.97          & 59.67          & 71.22          \\
\multicolumn{1}{c|}{}                            &                                 & \multicolumn{1}{c|}{Early Fusion}     & \textbf{93.56} & 14.2                              & \textbf{24.77} & \textbf{25.84} & \textbf{36.19} & \textbf{65.21} & \textbf{79.83} \\ \cline{2-10} 
\multicolumn{1}{c|}{}                            & \multirow{4}{*}{Medium}         & \multicolumn{1}{c|}{Late Fusion Rand} & 92.63          & 9.47                              & 17.62          & 20.61          & 28.18          & 57.04          & 71.42          \\
\multicolumn{1}{c|}{}                            &                                 & \multicolumn{1}{c|}{Late Fusion Max}  & 93.62          & 9.19                              & 17.16          & 20.42          & 28.24          & 59.32          & 73.80          \\
\multicolumn{1}{c|}{}                            &                                 & \multicolumn{1}{c|}{Late Fusion Mean} & 90.97          & 9.54                              & 17.33          & 20.42          & 29.31          & 58.15          & 71.00          \\
\multicolumn{1}{c|}{}                            &                                 & \multicolumn{1}{c|}{Early Fusion}     & \textbf{94.03} & \textbf{11.47}                    & \textbf{19.18} & \textbf{23.14} & \textbf{30.04} & \textbf{63.89} & \textbf{78.69} \\ \cline{2-10} 
\multicolumn{1}{c|}{}                            & \multirow{4}{*}{Tail}           & \multicolumn{1}{c|}{Late Fusion Rand} & 87.61          & 4.85                              & 8.66           & 13.26          & 8.58           & 24.25          & 41.39          \\
\multicolumn{1}{c|}{}                            &                                 & \multicolumn{1}{c|}{Late Fusion Max}  & 88.55          & 4.34                              & 8.03           & 12.96          & 9.13           & \textbf{27.46} & 41.43          \\
\multicolumn{1}{c|}{}                            &                                 & \multicolumn{1}{c|}{Late Fusion Mean} & 84.50          & 4.35                              & 8.09           & 12.80          & 9.38           & 25.96          & \textbf{42.57} \\
\multicolumn{1}{c|}{}                            &                                 & \multicolumn{1}{c|}{Early Fusion}     & \textbf{90.01} & \multicolumn{1}{l}{\textbf{8.97}} & \textbf{9.35}  & \textbf{13.30} & \textbf{10.04} & 27.02          & 42.32          \\ \hline
\end{tabular}
\end{table}

\begin{table}[t]
  \centering
  \tiny
   \resizebox{\linewidth}{!}{
\begin{tabular}{ccccccc}
\hline
Input Type &
  Diagnostic Granularity &
  3D Image Depth &
  Visual BackBone &
  Loss Fuction &
  Metrics &
  head/body/tail \\ \hline
\multicolumn{1}{c|}{\multirow{72}{*}{Single Image}} &
  \multicolumn{1}{c|}{\multirow{72}{*}{\begin{tabular}[c]{@{}c@{}}ICD-10 CM\\ (917 Classes)\end{tabular}}} &
  \multicolumn{1}{c|}{\multirow{36}{*}{16}} &
  \multicolumn{1}{c|}{\multirow{18}{*}{\begin{tabular}[c]{@{}c@{}}Resnet\\ (Resnet18 \& Resnet34)\end{tabular}}} &
  \multicolumn{1}{c|}{\multirow{9}{*}{BCE}} &
  AUC &
  87.99/87.98/84.62 \\
\multicolumn{1}{c|}{} &
  \multicolumn{1}{c|}{} &
  \multicolumn{1}{c|}{} &
  \multicolumn{1}{c|}{} &
  \multicolumn{1}{c|}{} &
  AP &
  11.44/6.86/3.56 \\
\multicolumn{1}{c|}{} &
  \multicolumn{1}{c|}{} &
  \multicolumn{1}{c|}{} &
  \multicolumn{1}{c|}{} &
  \multicolumn{1}{c|}{} &
  F1 &
  19.57/14.14/6.77 \\
\multicolumn{1}{c|}{} &
  \multicolumn{1}{c|}{} &
  \multicolumn{1}{c|}{} &
  \multicolumn{1}{c|}{} &
  \multicolumn{1}{c|}{} &
  MCC &
  20.42/16.73/10.58 \\
\multicolumn{1}{c|}{} &
  \multicolumn{1}{c|}{} &
  \multicolumn{1}{c|}{} &
  \multicolumn{1}{c|}{} &
  \multicolumn{1}{c|}{} &
  R@F0.01 &
  22.21/21.33/7.41 \\
\multicolumn{1}{c|}{} &
  \multicolumn{1}{c|}{} &
  \multicolumn{1}{c|}{} &
  \multicolumn{1}{c|}{} &
  \multicolumn{1}{c|}{} &
  R@F0.05 &
  48.16/47.18/20.86 \\
\multicolumn{1}{c|}{} &
  \multicolumn{1}{c|}{} &
  \multicolumn{1}{c|}{} &
  \multicolumn{1}{c|}{} &
  \multicolumn{1}{c|}{} &
  R@F0.1 &
  64.27/60.77/35.49 \\
\multicolumn{1}{c|}{} &
  \multicolumn{1}{c|}{} &
  \multicolumn{1}{c|}{} &
  \multicolumn{1}{c|}{} &
  \multicolumn{1}{c|}{} &
  R@F0.2 &
  79.07/75.55/60.39 \\
\multicolumn{1}{c|}{} &
  \multicolumn{1}{c|}{} &
  \multicolumn{1}{c|}{} &
  \multicolumn{1}{c|}{} &
  \multicolumn{1}{c|}{} &
  R@F0.5 &
  93.64/92.03/89.23 \\ \cline{5-7} 
\multicolumn{1}{c|}{} &
  \multicolumn{1}{c|}{} &
  \multicolumn{1}{c|}{} &
  \multicolumn{1}{c|}{} &
  \multicolumn{1}{c|}{\multirow{9}{*}{BCE Focal}} &
  AUC &
  87.24/86.94/82.63 \\
\multicolumn{1}{c|}{} &
  \multicolumn{1}{c|}{} &
  \multicolumn{1}{c|}{} &
  \multicolumn{1}{c|}{} &
  \multicolumn{1}{c|}{} &
  AP &
  11.23/7.16/4.81 \\
\multicolumn{1}{c|}{} &
  \multicolumn{1}{c|}{} &
  \multicolumn{1}{c|}{} &
  \multicolumn{1}{c|}{} &
  \multicolumn{1}{c|}{} &
  F1 &
  19.07/14.77/8.49 \\
\multicolumn{1}{c|}{} &
  \multicolumn{1}{c|}{} &
  \multicolumn{1}{c|}{} &
  \multicolumn{1}{c|}{} &
  \multicolumn{1}{c|}{} &
  MCC &
  20.27/17.31/12.18 \\
\multicolumn{1}{c|}{} &
  \multicolumn{1}{c|}{} &
  \multicolumn{1}{c|}{} &
  \multicolumn{1}{c|}{} &
  \multicolumn{1}{c|}{} &
  R@F0.01 &
  22.37/20.25/7.76 \\
\multicolumn{1}{c|}{} &
  \multicolumn{1}{c|}{} &
  \multicolumn{1}{c|}{} &
  \multicolumn{1}{c|}{} &
  \multicolumn{1}{c|}{} &
  R@F0.05 &
  46.80/45.10/21.86 \\
\multicolumn{1}{c|}{} &
  \multicolumn{1}{c|}{} &
  \multicolumn{1}{c|}{} &
  \multicolumn{1}{c|}{} &
  \multicolumn{1}{c|}{} &
  R@F0.1 &
  62.00/58.76/35.03 \\
\multicolumn{1}{c|}{} &
  \multicolumn{1}{c|}{} &
  \multicolumn{1}{c|}{} &
  \multicolumn{1}{c|}{} &
  \multicolumn{1}{c|}{} &
  R@F0.2 &
  76.03/72.92/58.97 \\
\multicolumn{1}{c|}{} &
  \multicolumn{1}{c|}{} &
  \multicolumn{1}{c|}{} &
  \multicolumn{1}{c|}{} &
  \multicolumn{1}{c|}{} &
  R@F0.5 &
  90.79/89.82/85.65 \\ \cline{4-7} 
\multicolumn{1}{c|}{} &
  \multicolumn{1}{c|}{} &
  \multicolumn{1}{c|}{} &
  \multicolumn{1}{c|}{\multirow{18}{*}{Vision Transformer}} &
  \multicolumn{1}{c|}{\multirow{9}{*}{BCE}} &
  AUC &
  84.76/83.61/78.74 \\
\multicolumn{1}{c|}{} &
  \multicolumn{1}{c|}{} &
  \multicolumn{1}{c|}{} &
  \multicolumn{1}{c|}{} &
  \multicolumn{1}{c|}{} &
  AP &
  7.21/3.53/1.54 \\
\multicolumn{1}{c|}{} &
  \multicolumn{1}{c|}{} &
  \multicolumn{1}{c|}{} &
  \multicolumn{1}{c|}{} &
  \multicolumn{1}{c|}{} &
  F1 &
  14.20/9.13/3.75 \\
\multicolumn{1}{c|}{} &
  \multicolumn{1}{c|}{} &
  \multicolumn{1}{c|}{} &
  \multicolumn{1}{c|}{} &
  \multicolumn{1}{c|}{} &
  MCC &
  15.49/11.36/6.97 \\
\multicolumn{1}{c|}{} &
  \multicolumn{1}{c|}{} &
  \multicolumn{1}{c|}{} &
  \multicolumn{1}{c|}{} &
  \multicolumn{1}{c|}{} &
  R@F0.01 &
  15.08/12.15/3.43 \\
\multicolumn{1}{c|}{} &
  \multicolumn{1}{c|}{} &
  \multicolumn{1}{c|}{} &
  \multicolumn{1}{c|}{} &
  \multicolumn{1}{c|}{} &
  R@F0.05 &
  39.26/33.43/13.38 \\
\multicolumn{1}{c|}{} &
  \multicolumn{1}{c|}{} &
  \multicolumn{1}{c|}{} &
  \multicolumn{1}{c|}{} &
  \multicolumn{1}{c|}{} &
  R@F0.1 &
  55.17/48.45/24.85 \\
\multicolumn{1}{c|}{} &
  \multicolumn{1}{c|}{} &
  \multicolumn{1}{c|}{} &
  \multicolumn{1}{c|}{} &
  \multicolumn{1}{c|}{} &
  R@F0.2 &
  72.70/65.89/45.04 \\
\multicolumn{1}{c|}{} &
  \multicolumn{1}{c|}{} &
  \multicolumn{1}{c|}{} &
  \multicolumn{1}{c|}{} &
  \multicolumn{1}{c|}{} &
  R@F0.5 &
  91.80/89.16/81.61 \\ \cline{5-7} 
\multicolumn{1}{c|}{} &
  \multicolumn{1}{c|}{} &
  \multicolumn{1}{c|}{} &
  \multicolumn{1}{c|}{} &
  \multicolumn{1}{c|}{\multirow{9}{*}{BCE Focal}} &
  AUC &
  83.51/82.05/79.12 \\
\multicolumn{1}{c|}{} &
  \multicolumn{1}{c|}{} &
  \multicolumn{1}{c|}{} &
  \multicolumn{1}{c|}{} &
  \multicolumn{1}{c|}{} &
  AP &
  6.60/3.32/2.31 \\
\multicolumn{1}{c|}{} &
  \multicolumn{1}{c|}{} &
  \multicolumn{1}{c|}{} &
  \multicolumn{1}{c|}{} &
  \multicolumn{1}{c|}{} &
  F1 &
  13.46/8.79/4.23 \\
\multicolumn{1}{c|}{} &
  \multicolumn{1}{c|}{} &
  \multicolumn{1}{c|}{} &
  \multicolumn{1}{c|}{} &
  \multicolumn{1}{c|}{} &
  MCC &
  14.64/11.14/7.37 \\
\multicolumn{1}{c|}{} &
  \multicolumn{1}{c|}{} &
  \multicolumn{1}{c|}{} &
  \multicolumn{1}{c|}{} &
  \multicolumn{1}{c|}{} &
  R@F0.01 &
  13.79/11.52/3.65 \\
\multicolumn{1}{c|}{} &
  \multicolumn{1}{c|}{} &
  \multicolumn{1}{c|}{} &
  \multicolumn{1}{c|}{} &
  \multicolumn{1}{c|}{} &
  R@F0.05 &
  36.40/30.98/13.82 \\
\multicolumn{1}{c|}{} &
  \multicolumn{1}{c|}{} &
  \multicolumn{1}{c|}{} &
  \multicolumn{1}{c|}{} &
  \multicolumn{1}{c|}{} &
  R@F0.1 &
  51.65/44.66/27.39 \\
\multicolumn{1}{c|}{} &
  \multicolumn{1}{c|}{} &
  \multicolumn{1}{c|}{} &
  \multicolumn{1}{c|}{} &
  \multicolumn{1}{c|}{} &
  R@F0.2 &
  69.67/63.07/47.46 \\
\multicolumn{1}{c|}{} &
  \multicolumn{1}{c|}{} &
  \multicolumn{1}{c|}{} &
  \multicolumn{1}{c|}{} &
  \multicolumn{1}{c|}{} &
  R@F0.5 &
  90.38/86.38/81.49 \\ \cline{3-7} 
\multicolumn{1}{c|}{} &
  \multicolumn{1}{c|}{} &
  \multicolumn{1}{c|}{\multirow{36}{*}{32}} &
  \multicolumn{1}{c|}{\multirow{18}{*}{\begin{tabular}[c]{@{}c@{}}Resnet\\ (Resnet18 \& Resnet34)\end{tabular}}} &
  \multicolumn{1}{c|}{\multirow{9}{*}{BCE}} &
  AUC &
  89.12/88.98/86.41 \\
\multicolumn{1}{c|}{} &
  \multicolumn{1}{c|}{} &
  \multicolumn{1}{c|}{} &
  \multicolumn{1}{c|}{} &
  \multicolumn{1}{c|}{} &
  AP &
  12.39/7.72/4.46 \\
\multicolumn{1}{c|}{} &
  \multicolumn{1}{c|}{} &
  \multicolumn{1}{c|}{} &
  \multicolumn{1}{c|}{} &
  \multicolumn{1}{c|}{} &
  F1 &
  20.51/15.44/8.54 \\
\multicolumn{1}{c|}{} &
  \multicolumn{1}{c|}{} &
  \multicolumn{1}{c|}{} &
  \multicolumn{1}{c|}{} &
  \multicolumn{1}{c|}{} &
  MCC &
  21.72/18.02/12.48 \\
\multicolumn{1}{c|}{} &
  \multicolumn{1}{c|}{} &
  \multicolumn{1}{c|}{} &
  \multicolumn{1}{c|}{} &
  \multicolumn{1}{c|}{} &
  R@F0.01 &
  24.12/22.52/8.37 \\
\multicolumn{1}{c|}{} &
  \multicolumn{1}{c|}{} &
  \multicolumn{1}{c|}{} &
  \multicolumn{1}{c|}{} &
  \multicolumn{1}{c|}{} &
  R@F0.05 &
  51.82/50.50/23.74 \\
\multicolumn{1}{c|}{} &
  \multicolumn{1}{c|}{} &
  \multicolumn{1}{c|}{} &
  \multicolumn{1}{c|}{} &
  \multicolumn{1}{c|}{} &
  R@F0.1 &
  67.93/63.96/37.12 \\
\multicolumn{1}{c|}{} &
  \multicolumn{1}{c|}{} &
  \multicolumn{1}{c|}{} &
  \multicolumn{1}{c|}{} &
  \multicolumn{1}{c|}{} &
  R@F0.2 &
  82.16/77.98/67.44 \\
\multicolumn{1}{c|}{} &
  \multicolumn{1}{c|}{} &
  \multicolumn{1}{c|}{} &
  \multicolumn{1}{c|}{} &
  \multicolumn{1}{c|}{} &
  R@F0.5 &
  95.13/93.70/89.86 \\ \cline{5-7} 
\multicolumn{1}{c|}{} &
  \multicolumn{1}{c|}{} &
  \multicolumn{1}{c|}{} &
  \multicolumn{1}{c|}{} &
  \multicolumn{1}{c|}{\multirow{9}{*}{BCE Focal}} &
  AUC &
  87.36/86.74/82.62 \\
\multicolumn{1}{c|}{} &
  \multicolumn{1}{c|}{} &
  \multicolumn{1}{c|}{} &
  \multicolumn{1}{c|}{} &
  \multicolumn{1}{c|}{} &
  AP &
  11.06/7.22/4.63 \\
\multicolumn{1}{c|}{} &
  \multicolumn{1}{c|}{} &
  \multicolumn{1}{c|}{} &
  \multicolumn{1}{c|}{} &
  \multicolumn{1}{c|}{} &
  F1 &
  19.09/15.04/8.36 \\
\multicolumn{1}{c|}{} &
  \multicolumn{1}{c|}{} &
  \multicolumn{1}{c|}{} &
  \multicolumn{1}{c|}{} &
  \multicolumn{1}{c|}{} &
  MCC &
  19.99/17.40/12.03 \\
\multicolumn{1}{c|}{} &
  \multicolumn{1}{c|}{} &
  \multicolumn{1}{c|}{} &
  \multicolumn{1}{c|}{} &
  \multicolumn{1}{c|}{} &
  R@F0.01 &
  22.49/20.56/7.99 \\
\multicolumn{1}{c|}{} &
  \multicolumn{1}{c|}{} &
  \multicolumn{1}{c|}{} &
  \multicolumn{1}{c|}{} &
  \multicolumn{1}{c|}{} &
  R@F0.05 &
  48.13/43.61/20.04 \\
\multicolumn{1}{c|}{} &
  \multicolumn{1}{c|}{} &
  \multicolumn{1}{c|}{} &
  \multicolumn{1}{c|}{} &
  \multicolumn{1}{c|}{} &
  R@F0.1 &
  62.64/56.44/30.26 \\
\multicolumn{1}{c|}{} &
  \multicolumn{1}{c|}{} &
  \multicolumn{1}{c|}{} &
  \multicolumn{1}{c|}{} &
  \multicolumn{1}{c|}{} &
  R@F0.2 &
  76.7370.62/47.25 \\
\multicolumn{1}{c|}{} &
  \multicolumn{1}{c|}{} &
  \multicolumn{1}{c|}{} &
  \multicolumn{1}{c|}{} &
  \multicolumn{1}{c|}{} &
  R@F0.5 &
  92.04/88.37/80.15 \\ \cline{4-7} 
\multicolumn{1}{c|}{} &
  \multicolumn{1}{c|}{} &
  \multicolumn{1}{c|}{} &
  \multicolumn{1}{c|}{\multirow{18}{*}{Vision Transformer}} &
  \multicolumn{1}{c|}{\multirow{9}{*}{BCE}} &
  AUC &
  85.46/84.78/82.55 \\
\multicolumn{1}{c|}{} &
  \multicolumn{1}{c|}{} &
  \multicolumn{1}{c|}{} &
  \multicolumn{1}{c|}{} &
  \multicolumn{1}{c|}{} &
  AP &
  8.08/4.95/2.95 \\
\multicolumn{1}{c|}{} &
  \multicolumn{1}{c|}{} &
  \multicolumn{1}{c|}{} &
  \multicolumn{1}{c|}{} &
  \multicolumn{1}{c|}{} &
  F1 &
  15.29/11.10/5.77 \\
\multicolumn{1}{c|}{} &
  \multicolumn{1}{c|}{} &
  \multicolumn{1}{c|}{} &
  \multicolumn{1}{c|}{} &
  \multicolumn{1}{c|}{} &
  MCC &
  16.40/13.41/9.29 \\
\multicolumn{1}{c|}{} &
  \multicolumn{1}{c|}{} &
  \multicolumn{1}{c|}{} &
  \multicolumn{1}{c|}{} &
  \multicolumn{1}{c|}{} &
  R@F0.01 &
  16.50/14.59/5.45 \\
\multicolumn{1}{c|}{} &
  \multicolumn{1}{c|}{} &
  \multicolumn{1}{c|}{} &
  \multicolumn{1}{c|}{} &
  \multicolumn{1}{c|}{} &
  R@F0.05 &
  40.81/37.16/16.63 \\
\multicolumn{1}{c|}{} &
  \multicolumn{1}{c|}{} &
  \multicolumn{1}{c|}{} &
  \multicolumn{1}{c|}{} &
  \multicolumn{1}{c|}{} &
  R@F0.1 &
  57.20/52.30/30.35 \\
\multicolumn{1}{c|}{} &
  \multicolumn{1}{c|}{} &
  \multicolumn{1}{c|}{} &
  \multicolumn{1}{c|}{} &
  \multicolumn{1}{c|}{} &
  R@F0.2 &
  74.35/68.78/54.96 \\
\multicolumn{1}{c|}{} &
  \multicolumn{1}{c|}{} &
  \multicolumn{1}{c|}{} &
  \multicolumn{1}{c|}{} &
  \multicolumn{1}{c|}{} &
  R@F0.5 &
  92.18/90.55/84.09 \\ \cline{5-7} 
\multicolumn{1}{c|}{} &
  \multicolumn{1}{c|}{} &
  \multicolumn{1}{c|}{} &
  \multicolumn{1}{c|}{} &
  \multicolumn{1}{c|}{\multirow{9}{*}{BCE Focal}} &
  AUC &
  84.80/83.63/79.24 \\
\multicolumn{1}{c|}{} &
  \multicolumn{1}{c|}{} &
  \multicolumn{1}{c|}{} &
  \multicolumn{1}{c|}{} &
  \multicolumn{1}{c|}{} &
  AP &
  7.04/4.12/2.07 \\
\multicolumn{1}{c|}{} &
  \multicolumn{1}{c|}{} &
  \multicolumn{1}{c|}{} &
  \multicolumn{1}{c|}{} &
  \multicolumn{1}{c|}{} &
  F1 &
  13.91/10.12/4.72 \\
\multicolumn{1}{c|}{} &
  \multicolumn{1}{c|}{} &
  \multicolumn{1}{c|}{} &
  \multicolumn{1}{c|}{} &
  \multicolumn{1}{c|}{} &
  MCC &
  15.07/12.38/8.21 \\
\multicolumn{1}{c|}{} &
  \multicolumn{1}{c|}{} &
  \multicolumn{1}{c|}{} &
  \multicolumn{1}{c|}{} &
  \multicolumn{1}{c|}{} &
  R@F0.01 &
  14.31/12.46/4.72 \\
\multicolumn{1}{c|}{} &
  \multicolumn{1}{c|}{} &
  \multicolumn{1}{c|}{} &
  \multicolumn{1}{c|}{} &
  \multicolumn{1}{c|}{} &
  R@F0.05 &
  38.93/34.72/14.65 \\
\multicolumn{1}{c|}{} &
  \multicolumn{1}{c|}{} &
  \multicolumn{1}{c|}{} &
  \multicolumn{1}{c|}{} &
  \multicolumn{1}{c|}{} &
  R@F0.1 &
  54.67/48.92/26.28 \\
\multicolumn{1}{c|}{} &
  \multicolumn{1}{c|}{} &
  \multicolumn{1}{c|}{} &
  \multicolumn{1}{c|}{} &
  \multicolumn{1}{c|}{} &
  R@F0.2 &
  72.37/66.32/48.34 \\
\multicolumn{1}{c|}{} &
  \multicolumn{1}{c|}{} &
  \multicolumn{1}{c|}{} &
  \multicolumn{1}{c|}{} &
  \multicolumn{1}{c|}{} &
  R@F0.5 &
  91.76/88.96/81.62 \\ \hline
\end{tabular}
  }
  \caption{Performance on single-image input. Classified on ICD-10 CM Level}
  \vspace{-2mm}
\end{table}

\begin{table}[t]
  \centering
  \tiny
   \resizebox{\linewidth}{!}{
\begin{tabular}{cccccccc}
\hline
Input Format &
  Level &
  Image Depth &
  BackBone &
  Loss &
  Metrics &
  head/body/tail &
  Solo \\ \hline
\multicolumn{1}{c|}{\multirow{72}{*}{Single Image}} &
  \multicolumn{1}{c|}{\multirow{72}{*}{\begin{tabular}[c]{@{}c@{}}Specific Disease\\ (5569 Classes)\end{tabular}}} &
  \multicolumn{1}{c|}{\multirow{36}{*}{16}} &
  \multicolumn{1}{c|}{\multirow{18}{*}{\begin{tabular}[c]{@{}c@{}}Resnet\\ (Resnet18 \& Resnet34)\end{tabular}}} &
  \multicolumn{1}{c|}{\multirow{9}{*}{BCE}} &
  AUC &
  90.74/90.82/87.16 &
  88.55 \\
\multicolumn{1}{c|}{} &
  \multicolumn{1}{c|}{} &
  \multicolumn{1}{c|}{} &
  \multicolumn{1}{c|}{} &
  \multicolumn{1}{c|}{} &
  AP &
  11.71/8.08/3.83 &
  23.39 \\
\multicolumn{1}{c|}{} &
  \multicolumn{1}{c|}{} &
  \multicolumn{1}{c|}{} &
  \multicolumn{1}{c|}{} &
  \multicolumn{1}{c|}{} &
  F1 &
  20.07/15.70/7.31 &
  31.98 \\
\multicolumn{1}{c|}{} &
  \multicolumn{1}{c|}{} &
  \multicolumn{1}{c|}{} &
  \multicolumn{1}{c|}{} &
  \multicolumn{1}{c|}{} &
  MCC &
  21.45/18.14/11.99 &
  32.38 \\
\multicolumn{1}{c|}{} &
  \multicolumn{1}{c|}{} &
  \multicolumn{1}{c|}{} &
  \multicolumn{1}{c|}{} &
  \multicolumn{1}{c|}{} &
  R@F0.01 &
  27.59/24.11/7.92 &
  25.81 \\
\multicolumn{1}{c|}{} &
  \multicolumn{1}{c|}{} &
  \multicolumn{1}{c|}{} &
  \multicolumn{1}{c|}{} &
  \multicolumn{1}{c|}{} &
  R@F0.05 &
  56.36/53.93/20.45 &
  51.23 \\
\multicolumn{1}{c|}{} &
  \multicolumn{1}{c|}{} &
  \multicolumn{1}{c|}{} &
  \multicolumn{1}{c|}{} &
  \multicolumn{1}{c|}{} &
  R@F0.1 &
  70.33/68.69/32.22 &
  65.51 \\
\multicolumn{1}{c|}{} &
  \multicolumn{1}{c|}{} &
  \multicolumn{1}{c|}{} &
  \multicolumn{1}{c|}{} &
  \multicolumn{1}{c|}{} &
  R@F0.2 &
  84.10/82.78/54.48 &
  79.12 \\
\multicolumn{1}{c|}{} &
  \multicolumn{1}{c|}{} &
  \multicolumn{1}{c|}{} &
  \multicolumn{1}{c|}{} &
  \multicolumn{1}{c|}{} &
  R@F0.5 &
  95.70/95.46/87.12 &
  93.48 \\ \cline{5-8} 
\multicolumn{1}{c|}{} &
  \multicolumn{1}{c|}{} &
  \multicolumn{1}{c|}{} &
  \multicolumn{1}{c|}{} &
  \multicolumn{1}{c|}{\multirow{9}{*}{BCE Focal}} &
  AUC &
  90.60/92.09/86.63 &
  88.89 \\
\multicolumn{1}{c|}{} &
  \multicolumn{1}{c|}{} &
  \multicolumn{1}{c|}{} &
  \multicolumn{1}{c|}{} &
  \multicolumn{1}{c|}{} &
  AP &
  12.07/9.21/4.56 &
  23.71 \\
\multicolumn{1}{c|}{} &
  \multicolumn{1}{c|}{} &
  \multicolumn{1}{c|}{} &
  \multicolumn{1}{c|}{} &
  \multicolumn{1}{c|}{} &
  F1 &
  21.24/11.08/8.26 &
  32.02 \\
\multicolumn{1}{c|}{} &
  \multicolumn{1}{c|}{} &
  \multicolumn{1}{c|}{} &
  \multicolumn{1}{c|}{} &
  \multicolumn{1}{c|}{} &
  MCC &
  22.59/20.31/12.73 &
  32.84 \\
\multicolumn{1}{c|}{} &
  \multicolumn{1}{c|}{} &
  \multicolumn{1}{c|}{} &
  \multicolumn{1}{c|}{} &
  \multicolumn{1}{c|}{} &
  R@F0.01 &
  28.96/25.80/7.76 &
  25.93 \\
\multicolumn{1}{c|}{} &
  \multicolumn{1}{c|}{} &
  \multicolumn{1}{c|}{} &
  \multicolumn{1}{c|}{} &
  \multicolumn{1}{c|}{} &
  R@F0.05 &
  56.94/54.28/20.24 &
  51.40 \\
\multicolumn{1}{c|}{} &
  \multicolumn{1}{c|}{} &
  \multicolumn{1}{c|}{} &
  \multicolumn{1}{c|}{} &
  \multicolumn{1}{c|}{} &
  R@F0.1 &
  71.29/67.54/31.20 &
  66.38 \\
\multicolumn{1}{c|}{} &
  \multicolumn{1}{c|}{} &
  \multicolumn{1}{c|}{} &
  \multicolumn{1}{c|}{} &
  \multicolumn{1}{c|}{} &
  R@F0.2 &
  84.00/80.96/51.20 &
  79.69 \\
\multicolumn{1}{c|}{} &
  \multicolumn{1}{c|}{} &
  \multicolumn{1}{c|}{} &
  \multicolumn{1}{c|}{} &
  \multicolumn{1}{c|}{} &
  R@F0.5 &
  95.91/95.56/85.32 &
  92.82 \\ \cline{4-8} 
\multicolumn{1}{c|}{} &
  \multicolumn{1}{c|}{} &
  \multicolumn{1}{c|}{} &
  \multicolumn{1}{c|}{\multirow{18}{*}{Vision Transformer}} &
  \multicolumn{1}{c|}{\multirow{9}{*}{BCE}} &
  AUC &
  87.23/87.40/87.89 &
  83.12 \\
\multicolumn{1}{c|}{} &
  \multicolumn{1}{c|}{} &
  \multicolumn{1}{c|}{} &
  \multicolumn{1}{c|}{} &
  \multicolumn{1}{c|}{} &
  AP &
  6.61/4.47/1.93 &
  14.31 \\
\multicolumn{1}{c|}{} &
  \multicolumn{1}{c|}{} &
  \multicolumn{1}{c|}{} &
  \multicolumn{1}{c|}{} &
  \multicolumn{1}{c|}{} &
  F1 &
  14.17/10.15/4.16 &
  22.61 \\
\multicolumn{1}{c|}{} &
  \multicolumn{1}{c|}{} &
  \multicolumn{1}{c|}{} &
  \multicolumn{1}{c|}{} &
  \multicolumn{1}{c|}{} &
  MCC &
  15.68/13.13/8.09 &
  23.36 \\
\multicolumn{1}{c|}{} &
  \multicolumn{1}{c|}{} &
  \multicolumn{1}{c|}{} &
  \multicolumn{1}{c|}{} &
  \multicolumn{1}{c|}{} &
  R@F0.01 &
  18.36/14.27/3.77 &
  15.30 \\
\multicolumn{1}{c|}{} &
  \multicolumn{1}{c|}{} &
  \multicolumn{1}{c|}{} &
  \multicolumn{1}{c|}{} &
  \multicolumn{1}{c|}{} &
  R@F0.05 &
  44.23/40.18/15.03 &
  37.67 \\
\multicolumn{1}{c|}{} &
  \multicolumn{1}{c|}{} &
  \multicolumn{1}{c|}{} &
  \multicolumn{1}{c|}{} &
  \multicolumn{1}{c|}{} &
  R@F0.1 &
  58.36/56.96/27.39 &
  51.98 \\
\multicolumn{1}{c|}{} &
  \multicolumn{1}{c|}{} &
  \multicolumn{1}{c|}{} &
  \multicolumn{1}{c|}{} &
  \multicolumn{1}{c|}{} &
  R@F0.2 &
  74.17/74.47/50.30 &
  68.50 \\
\multicolumn{1}{c|}{} &
  \multicolumn{1}{c|}{} &
  \multicolumn{1}{c|}{} &
  \multicolumn{1}{c|}{} &
  \multicolumn{1}{c|}{} &
  R@F0.5 &
  93.91/94.32/87.22 &
  80.16 \\ \cline{5-8} 
\multicolumn{1}{c|}{} &
  \multicolumn{1}{c|}{} &
  \multicolumn{1}{c|}{} &
  \multicolumn{1}{c|}{} &
  \multicolumn{1}{c|}{\multirow{9}{*}{BCE Focal}} &
  AUC &
  86.83/86.88/81.53 &
  83.15 \\
\multicolumn{1}{c|}{} &
  \multicolumn{1}{c|}{} &
  \multicolumn{1}{c|}{} &
  \multicolumn{1}{c|}{} &
  \multicolumn{1}{c|}{} &
  AP &
  6.13/4.10/1.93 &
  13.25 \\
\multicolumn{1}{c|}{} &
  \multicolumn{1}{c|}{} &
  \multicolumn{1}{c|}{} &
  \multicolumn{1}{c|}{} &
  \multicolumn{1}{c|}{} &
  F1 &
  13.53/9.79/4.06 &
  22.07 \\
\multicolumn{1}{c|}{} &
  \multicolumn{1}{c|}{} &
  \multicolumn{1}{c|}{} &
  \multicolumn{1}{c|}{} &
  \multicolumn{1}{c|}{} &
  MCC &
  15.01/12.51/7.75 &
  22.86 \\
\multicolumn{1}{c|}{} &
  \multicolumn{1}{c|}{} &
  \multicolumn{1}{c|}{} &
  \multicolumn{1}{c|}{} &
  \multicolumn{1}{c|}{} &
  R@F0.01 &
  17.69/13.67/3.28 &
  14.56 \\
\multicolumn{1}{c|}{} &
  \multicolumn{1}{c|}{} &
  \multicolumn{1}{c|}{} &
  \multicolumn{1}{c|}{} &
  \multicolumn{1}{c|}{} &
  R@F0.05 &
  44.77/38.40/13.94 &
  38.01 \\
\multicolumn{1}{c|}{} &
  \multicolumn{1}{c|}{} &
  \multicolumn{1}{c|}{} &
  \multicolumn{1}{c|}{} &
  \multicolumn{1}{c|}{} &
  R@F0.1 &
  59.75/54.52/26.69 &
  52.34 \\
\multicolumn{1}{c|}{} &
  \multicolumn{1}{c|}{} &
  \multicolumn{1}{c|}{} &
  \multicolumn{1}{c|}{} &
  \multicolumn{1}{c|}{} &
  R@F0.2 &
  75.88/73.08/84.27 &
  69.07 \\
\multicolumn{1}{c|}{} &
  \multicolumn{1}{c|}{} &
  \multicolumn{1}{c|}{} &
  \multicolumn{1}{c|}{} &
  \multicolumn{1}{c|}{} &
  R@F0.5 &
  93.22/93.50/84.27 &
  88.71 \\ \cline{3-8} 
\multicolumn{1}{c|}{} &
  \multicolumn{1}{c|}{} &
  \multicolumn{1}{c|}{\multirow{36}{*}{32}} &
  \multicolumn{1}{c|}{\multirow{18}{*}{\begin{tabular}[c]{@{}c@{}}Resnet\\ (Resnet18 \& Resnet34)\end{tabular}}} &
  \multicolumn{1}{c|}{\multirow{9}{*}{BCE}} &
  AUC &
  91.32/92.63/87.61 &
  88.84 \\
\multicolumn{1}{c|}{} &
  \multicolumn{1}{c|}{} &
  \multicolumn{1}{c|}{} &
  \multicolumn{1}{c|}{} &
  \multicolumn{1}{c|}{} &
  AP &
  14.08/9.47/4.85 &
  24.40 \\
\multicolumn{1}{c|}{} &
  \multicolumn{1}{c|}{} &
  \multicolumn{1}{c|}{} &
  \multicolumn{1}{c|}{} &
  \multicolumn{1}{c|}{} &
  F1 &
  22.93/17.62/8.66 &
  33.13 \\
\multicolumn{1}{c|}{} &
  \multicolumn{1}{c|}{} &
  \multicolumn{1}{c|}{} &
  \multicolumn{1}{c|}{} &
  \multicolumn{1}{c|}{} &
  MCC &
  24.61/20.61/13.26 &
  33.88 \\
\multicolumn{1}{c|}{} &
  \multicolumn{1}{c|}{} &
  \multicolumn{1}{c|}{} &
  \multicolumn{1}{c|}{} &
  \multicolumn{1}{c|}{} &
  R@F0.01 &
  30.68/28.18/8.58 &
  27.31 \\
\multicolumn{1}{c|}{} &
  \multicolumn{1}{c|}{} &
  \multicolumn{1}{c|}{} &
  \multicolumn{1}{c|}{} &
  \multicolumn{1}{c|}{} &
  R@F0.05 &
  58.78/57.04/24.25 &
  52.46 \\
\multicolumn{1}{c|}{} &
  \multicolumn{1}{c|}{} &
  \multicolumn{1}{c|}{} &
  \multicolumn{1}{c|}{} &
  \multicolumn{1}{c|}{} &
  R@F0.1 &
  73.79/71.42/41.39 &
  67.10 \\
\multicolumn{1}{c|}{} &
  \multicolumn{1}{c|}{} &
  \multicolumn{1}{c|}{} &
  \multicolumn{1}{c|}{} &
  \multicolumn{1}{c|}{} &
  R@F0.2 &
  85.33/85.06/66.42 &
  80.41 \\
\multicolumn{1}{c|}{} &
  \multicolumn{1}{c|}{} &
  \multicolumn{1}{c|}{} &
  \multicolumn{1}{c|}{} &
  \multicolumn{1}{c|}{} &
  R@F0.5 &
  96.12/97.41/91.54 &
  93.67 \\ \cline{5-8} 
\multicolumn{1}{c|}{} &
  \multicolumn{1}{c|}{} &
  \multicolumn{1}{c|}{} &
  \multicolumn{1}{c|}{} &
  \multicolumn{1}{c|}{\multirow{9}{*}{BCE Focal}} &
  AUC &
   &
   \\
\multicolumn{1}{c|}{} &
  \multicolumn{1}{c|}{} &
  \multicolumn{1}{c|}{} &
  \multicolumn{1}{c|}{} &
  \multicolumn{1}{c|}{} &
  AP &
   &
   \\
\multicolumn{1}{c|}{} &
  \multicolumn{1}{c|}{} &
  \multicolumn{1}{c|}{} &
  \multicolumn{1}{c|}{} &
  \multicolumn{1}{c|}{} &
  F1 &
   &
   \\
\multicolumn{1}{c|}{} &
  \multicolumn{1}{c|}{} &
  \multicolumn{1}{c|}{} &
  \multicolumn{1}{c|}{} &
  \multicolumn{1}{c|}{} &
  MCC &
   &
   \\
\multicolumn{1}{c|}{} &
  \multicolumn{1}{c|}{} &
  \multicolumn{1}{c|}{} &
  \multicolumn{1}{c|}{} &
  \multicolumn{1}{c|}{} &
  R@F0.01 &
   &
   \\
\multicolumn{1}{c|}{} &
  \multicolumn{1}{c|}{} &
  \multicolumn{1}{c|}{} &
  \multicolumn{1}{c|}{} &
  \multicolumn{1}{c|}{} &
  R@F0.05 &
   &
   \\
\multicolumn{1}{c|}{} &
  \multicolumn{1}{c|}{} &
  \multicolumn{1}{c|}{} &
  \multicolumn{1}{c|}{} &
  \multicolumn{1}{c|}{} &
  R@F0.1 &
   &
   \\
\multicolumn{1}{c|}{} &
  \multicolumn{1}{c|}{} &
  \multicolumn{1}{c|}{} &
  \multicolumn{1}{c|}{} &
  \multicolumn{1}{c|}{} &
  R@F0.2 &
   &
   \\
\multicolumn{1}{c|}{} &
  \multicolumn{1}{c|}{} &
  \multicolumn{1}{c|}{} &
  \multicolumn{1}{c|}{} &
  \multicolumn{1}{c|}{} &
  R@F0.5 &
   &
   \\ \cline{4-8} 
\multicolumn{1}{c|}{} &
  \multicolumn{1}{c|}{} &
  \multicolumn{1}{c|}{} &
  \multicolumn{1}{c|}{\multirow{18}{*}{Vision Transformer}} &
  \multicolumn{1}{c|}{\multirow{9}{*}{BCE}} &
  AUC &
  88.42/89.78/83.98 &
  84.44 \\
\multicolumn{1}{c|}{} &
  \multicolumn{1}{c|}{} &
  \multicolumn{1}{c|}{} &
  \multicolumn{1}{c|}{} &
  \multicolumn{1}{c|}{} &
  AP &
  8.14/5.43/2.52 &
  15.40 \\
\multicolumn{1}{c|}{} &
  \multicolumn{1}{c|}{} &
  \multicolumn{1}{c|}{} &
  \multicolumn{1}{c|}{} &
  \multicolumn{1}{c|}{} &
  F1 &
  16.17/11.65/5.17 &
  24.46 \\
\multicolumn{1}{c|}{} &
  \multicolumn{1}{c|}{} &
  \multicolumn{1}{c|}{} &
  \multicolumn{1}{c|}{} &
  \multicolumn{1}{c|}{} &
  MCC &
  17.84/14.86/9.28 &
  24.78 \\
\multicolumn{1}{c|}{} &
  \multicolumn{1}{c|}{} &
  \multicolumn{1}{c|}{} &
  \multicolumn{1}{c|}{} &
  \multicolumn{1}{c|}{} &
  R@F0.01 &
  21.85/17.43/5.01 &
  17.25 \\
\multicolumn{1}{c|}{} &
  \multicolumn{1}{c|}{} &
  \multicolumn{1}{c|}{} &
  \multicolumn{1}{c|}{} &
  \multicolumn{1}{c|}{} &
  R@F0.05 &
  49.14/44.71/20.57 &
  40.87 \\
\multicolumn{1}{c|}{} &
  \multicolumn{1}{c|}{} &
  \multicolumn{1}{c|}{} &
  \multicolumn{1}{c|}{} &
  \multicolumn{1}{c|}{} &
  R@F0.1 &
  65.10/61.30/37.20 &
  54.86 \\
\multicolumn{1}{c|}{} &
  \multicolumn{1}{c|}{} &
  \multicolumn{1}{c|}{} &
  \multicolumn{1}{c|}{} &
  \multicolumn{1}{c|}{} &
  R@F0.2 &
  80.33/79.04/61.31 &
  70.54 \\
\multicolumn{1}{c|}{} &
  \multicolumn{1}{c|}{} &
  \multicolumn{1}{c|}{} &
  \multicolumn{1}{c|}{} &
  \multicolumn{1}{c|}{} &
  R@F0.5 &
  94.20/95.07/87.65 &
  90.15 \\ \cline{5-8} 
\multicolumn{1}{c|}{} &
  \multicolumn{1}{c|}{} &
  \multicolumn{1}{c|}{} &
  \multicolumn{1}{c|}{} &
  \multicolumn{1}{c|}{\multirow{9}{*}{BCE Focal}} &
  AUC &
   &
   \\
\multicolumn{1}{c|}{} &
  \multicolumn{1}{c|}{} &
  \multicolumn{1}{c|}{} &
  \multicolumn{1}{c|}{} &
  \multicolumn{1}{c|}{} &
  AP &
   &
   \\
\multicolumn{1}{c|}{} &
  \multicolumn{1}{c|}{} &
  \multicolumn{1}{c|}{} &
  \multicolumn{1}{c|}{} &
  \multicolumn{1}{c|}{} &
  F1 &
   &
   \\
\multicolumn{1}{c|}{} &
  \multicolumn{1}{c|}{} &
  \multicolumn{1}{c|}{} &
  \multicolumn{1}{c|}{} &
  \multicolumn{1}{c|}{} &
  MCC &
   &
   \\
\multicolumn{1}{c|}{} &
  \multicolumn{1}{c|}{} &
  \multicolumn{1}{c|}{} &
  \multicolumn{1}{c|}{} &
  \multicolumn{1}{c|}{} &
  R@F0.01 &
   &
   \\
\multicolumn{1}{c|}{} &
  \multicolumn{1}{c|}{} &
  \multicolumn{1}{c|}{} &
  \multicolumn{1}{c|}{} &
  \multicolumn{1}{c|}{} &
  R@F0.05 &
   &
   \\
\multicolumn{1}{c|}{} &
  \multicolumn{1}{c|}{} &
  \multicolumn{1}{c|}{} &
  \multicolumn{1}{c|}{} &
  \multicolumn{1}{c|}{} &
  R@F0.1 &
   &
   \\
\multicolumn{1}{c|}{} &
  \multicolumn{1}{c|}{} &
  \multicolumn{1}{c|}{} &
  \multicolumn{1}{c|}{} &
  \multicolumn{1}{c|}{} &
  R@F0.2 &
   &
   \\
\multicolumn{1}{c|}{} &
  \multicolumn{1}{c|}{} &
  \multicolumn{1}{c|}{} &
  \multicolumn{1}{c|}{} &
  \multicolumn{1}{c|}{} &
  R@F0.5 &
   &
   \\ \hline
\end{tabular}
  }
  \caption{Performance on single-image input. Classified on Specific Disease Level}
  \vspace{-2mm}
\end{table}

\begin{table}[t]
  \centering
  \tiny
   \resizebox{\linewidth}{!}{
\begin{tabular}{ccccc}
\hline
Input Type                                                                                             & Diagnostic Granularity                                                                                           & Image Depth                              & Metrics & head/body/tail    \\ \hline
\multicolumn{1}{c|}{\multirow{36}{*}{\begin{tabular}[c]{@{}c@{}}Multi-mod\\ Multi Image\end{tabular}}} & \multicolumn{1}{c|}{\multirow{18}{*}{\begin{tabular}[c]{@{}c@{}}ICD-10 CM\\ (917 Classes)\end{tabular}}}         & \multicolumn{1}{c|}{\multirow{9}{*}{16}} & AUC     & 89.02/88.44/79.58 \\
\multicolumn{1}{c|}{}                                                                                  & \multicolumn{1}{c|}{}                                                                                            & \multicolumn{1}{c|}{}                    & AP      & 12.42/7.36/3.10   \\
\multicolumn{1}{c|}{}                                                                                  & \multicolumn{1}{c|}{}                                                                                            & \multicolumn{1}{c|}{}                    & F1      & 21.70/15.21/6.02  \\
\multicolumn{1}{c|}{}                                                                                  & \multicolumn{1}{c|}{}                                                                                            & \multicolumn{1}{c|}{}                    & MCC     & 22.93/18.12/9.31  \\
\multicolumn{1}{c|}{}                                                                                  & \multicolumn{1}{c|}{}                                                                                            & \multicolumn{1}{c|}{}                    & R@F0.01 & 24.54/23.22/6.45  \\
\multicolumn{1}{c|}{}                                                                                  & \multicolumn{1}{c|}{}                                                                                            & \multicolumn{1}{c|}{}                    & R@F0.05 & 53.36/48.44/16.79 \\
\multicolumn{1}{c|}{}                                                                                  & \multicolumn{1}{c|}{}                                                                                            & \multicolumn{1}{c|}{}                    & R@F0.1  & 68.06/61.83/28.03 \\
\multicolumn{1}{c|}{}                                                                                  & \multicolumn{1}{c|}{}                                                                                            & \multicolumn{1}{c|}{}                    & R@F0.2  & 81.51/75.91/51.10 \\
\multicolumn{1}{c|}{}                                                                                  & \multicolumn{1}{c|}{}                                                                                            & \multicolumn{1}{c|}{}                    & R@F0.5  & 94.01/93.72/81.28 \\ \cline{3-5} 
\multicolumn{1}{c|}{}                                                                                  & \multicolumn{1}{c|}{}                                                                                            & \multicolumn{1}{c|}{\multirow{9}{*}{32}} & AUC     & 91.15/89.92/83.87 \\
\multicolumn{1}{c|}{}                                                                                  & \multicolumn{1}{c|}{}                                                                                            & \multicolumn{1}{c|}{}                    & AP      & 14.38/9.24/4.27   \\
\multicolumn{1}{c|}{}                                                                                  & \multicolumn{1}{c|}{}                                                                                            & \multicolumn{1}{c|}{}                    & F1      & 22.83/17.14/7.88  \\
\multicolumn{1}{c|}{}                                                                                  & \multicolumn{1}{c|}{}                                                                                            & \multicolumn{1}{c|}{}                    & MCC     & 24.12/19.86/11.74 \\
\multicolumn{1}{c|}{}                                                                                  & \multicolumn{1}{c|}{}                                                                                            & \multicolumn{1}{c|}{}                    & R@F0.01 & 28.39/26.89/8.83  \\
\multicolumn{1}{c|}{}                                                                                  & \multicolumn{1}{c|}{}                                                                                            & \multicolumn{1}{c|}{}                    & R@F0.05 & 58.82/54.31/22.69 \\
\multicolumn{1}{c|}{}                                                                                  & \multicolumn{1}{c|}{}                                                                                            & \multicolumn{1}{c|}{}                    & R@F0.1  & 73.48/66.52/36.01 \\
\multicolumn{1}{c|}{}                                                                                  & \multicolumn{1}{c|}{}                                                                                            & \multicolumn{1}{c|}{}                    & R@F0.2  & 85.23/79.69/58.18 \\
\multicolumn{1}{c|}{}                                                                                  & \multicolumn{1}{c|}{}                                                                                            & \multicolumn{1}{c|}{}                    & R@F0.5  & 96.83/94.73/86.70 \\ \cline{2-5} 
\multicolumn{1}{c|}{}                                                                                  & \multicolumn{1}{c|}{\multirow{18}{*}{\begin{tabular}[c]{@{}c@{}}Specific Disease\\ (5569 Classes)\end{tabular}}} & \multicolumn{1}{c|}{\multirow{9}{*}{16}} & AUC     & 92.82/93.87/88.13 \\
\multicolumn{1}{c|}{}                                                                                  & \multicolumn{1}{c|}{}                                                                                            & \multicolumn{1}{c|}{}                    & AP      & 14.09/10.01/4.43  \\
\multicolumn{1}{c|}{}                                                                                  & \multicolumn{1}{c|}{}                                                                                            & \multicolumn{1}{c|}{}                    & F1      & 23.41/17.88/8.12  \\
\multicolumn{1}{c|}{}                                                                                  & \multicolumn{1}{c|}{}                                                                                            & \multicolumn{1}{c|}{}                    & MCC     & 24.69/21.00/12.84 \\
\multicolumn{1}{c|}{}                                                                                  & \multicolumn{1}{c|}{}                                                                                            & \multicolumn{1}{c|}{}                    & R@F0.01 & 33.88/27.75/9.04  \\
\multicolumn{1}{c|}{}                                                                                  & \multicolumn{1}{c|}{}                                                                                            & \multicolumn{1}{c|}{}                    & R@F0.05 & 63.34/61.15/24.48 \\
\multicolumn{1}{c|}{}                                                                                  & \multicolumn{1}{c|}{}                                                                                            & \multicolumn{1}{c|}{}                    & R@F0.1  & 76.22/73.54/39.32 \\
\multicolumn{1}{c|}{}                                                                                  & \multicolumn{1}{c|}{}                                                                                            & \multicolumn{1}{c|}{}                    & R@F0.2  & 87.54/86.79/66.79 \\
\multicolumn{1}{c|}{}                                                                                  & \multicolumn{1}{c|}{}                                                                                            & \multicolumn{1}{c|}{}                    & R@F0.5  & 97.28/97.69/92.08 \\ \cline{3-5} 
\multicolumn{1}{c|}{}                                                                                  & \multicolumn{1}{c|}{}                                                                                            & \multicolumn{1}{c|}{\multirow{9}{*}{32}} & AUC     &                   \\
\multicolumn{1}{c|}{}                                                                                  & \multicolumn{1}{c|}{}                                                                                            & \multicolumn{1}{c|}{}                    & AP      &                   \\
\multicolumn{1}{c|}{}                                                                                  & \multicolumn{1}{c|}{}                                                                                            & \multicolumn{1}{c|}{}                    & F1      &                   \\
\multicolumn{1}{c|}{}                                                                                  & \multicolumn{1}{c|}{}                                                                                            & \multicolumn{1}{c|}{}                    & MCC     &                   \\
\multicolumn{1}{c|}{}                                                                                  & \multicolumn{1}{c|}{}                                                                                            & \multicolumn{1}{c|}{}                    & R@F0.01 &                   \\
\multicolumn{1}{c|}{}                                                                                  & \multicolumn{1}{c|}{}                                                                                            & \multicolumn{1}{c|}{}                    & R@F0.05 &                   \\
\multicolumn{1}{c|}{}                                                                                  & \multicolumn{1}{c|}{}                                                                                            & \multicolumn{1}{c|}{}                    & R@F0.1  &                   \\
\multicolumn{1}{c|}{}                                                                                  & \multicolumn{1}{c|}{}                                                                                            & \multicolumn{1}{c|}{}                    & R@F0.2  &                   \\
\multicolumn{1}{c|}{}                                                                                  & \multicolumn{1}{c|}{}                                                                                            & \multicolumn{1}{c|}{}                    & R@F0.5  &                   \\ \hline
\end{tabular}
  }
  \caption{Performance on multi-image input. Classified on both Disease and ICD Level}
  \vspace{-2mm}
\end{table}